\documentclass[a4paper,fleqn]{cas-sc}

\usepackage[numbers]{natbib}
\usepackage{caption}
\usepackage{subcaption}
\usepackage{float} 
\newtheorem{remark}{Remark}
\def\tsc#1{\csdef{#1}{\textsc{\lowercase{#1}}\xspace}}
\tsc{WGM}
\tsc{QE}
\tsc{EP}
\tsc{PMS}
\tsc{BEC}
\tsc{DE}

\begin{document}
\let\WriteBookmarks\relax
\def\floatpagepagefraction{1}
\def\textpagefraction{.001}
\shorttitle{A Deep Learning based Detection Method for Combined Integrity-Availability Cyber Attacks in Power System}
\shortauthors{W. Xu, F. Teng}

\title [mode = title]{A Deep Learning based Detection Method for Combined Integrity-Availability Cyber Attacks in Power System}                      

\author[1]{Wangkun Xu}[orcid=0000-0001-9811-2673,
                         style=english]
\ead{wangkun.xu18@imperial.ac.uk}
\credit{Conceptualization of this study, Methodology, Software, Writing - Original Draft}
\address[1]{Control and Power Group, Department of Electrical and Electronic Engineering, Imperial College London, London, UK}


\author[1]{Fei Teng}[style=chinese, orcid = 0000-0002-6828-0294]
\ead{f.teng@imperial.ac.uk}
\cormark[1]
\credit{Conceptualization of this study, Supervision, Writing - Review \& Editing}

\cortext[cor1]{Corresponding author. Tel.: +44 (0)20 7594 6178. Address: Room 1113, Electrical and Electronic Engineering, South Kensington Campus, Imperial College London, SW7 2AZ, UK }

\begin{abstract}
As one of the largest and most complex systems on earth, power grid operation and control have stepped forward as a compound analysis on both physical and cyber layers which makes it vulnerable to assaults from economic and security considerations. In order to accomplish reliable power system control, it is necessary to do state estimation and bad data detection to remove the erroneous measurement. A new type of attack, namely the combined data integrity-availability attack, has recently been proposed, where the attackers can simultaneously manipulate and blind some measurements on the supervisory control and data acquisition system to mislead the control operation and keep stealthy. Compared with traditional false data injection attacks, this combined attack can further complicate and vitiate the model-based detection mechanism. To detect such attack, 
this paper proposes a novel random denoising long short-term memory autoencoder framework, where 
the spatial-temporal correlations of measurements can be explicitly captured and the unavailable data is countered by the random dropout layer. 
The proposed algorithm is evaluated and the performance is verified on a 118-bus system under various unseen attack attempts. 


\end{abstract}



\maketitle

\section{Introduction}

\subsection{Background}

The emerging application of information and communication techniques (ICT) on power system automation, monitoring, and control has reformed the modern power gird into a complex cyber-physical system (CPS) \cite{aslani2022novel}. Many flexible and artificial solutions are proposed to embrace the trend of electrification and digitalization at all cyber, physical, network, and communication layers to allow two-way communication between facilities and customers \cite{ibrahim2020machine, teng2016benefits}. However, this new opportunity also presents challenges in the safe and resilient operation of the cyber-physical power system under cyberattacks \cite{yohanandhan2022holistic}. Several cyber-physical attacks have been reported causing severe economic and human life losses, among which the incidence taking place in Ukraine may be the most recognizable one. In December 2015, more than 200k customers were influenced by the half-day grid blackout in Kiev. The follow-up investigation revealed that the supervisory control and data acquisition (SCADA) protocol was affected by hackers \cite{lee16analysis}.

Cyber-physical attacks can be classified according to their unique target and delivery methodologies \cite{musleh19survey}, such as denial-of-service (DoS) attacks in the network and communication layers where information flow packets are jammed or lost. In particular, most of the recent research is focused on the False Data Injection Attacks (FDIAs) which can be implemented on all four layers in cyber-physical power systems. Although FDIAs have been researched in many directions, such as DC microgrid operation \cite{bi2022false, tan2022false}, energy market\cite{gao2022novel}, frequency regulation \cite{chen2020load}, and voltage regulation \cite{choeum2021trilevel}, this paper mainly discusses FDIAs targetting on state estimation (SE). In principle, the control center retrieves the operational states of individual buses by the measurements observed from the remote terminal unit (RTU) and/or phasor measurement unit (PMU) in every few seconds to minutes. The estimated states are used in energy management systems (EMS) for contingency analysis, automatic generation control (AGC), load forecasting, etc. \cite{abur04power}. Therefore, FDIA is defined as a direct manipulation on the measurements with the purpose of deviating the estimated state that can mislead the EMS's economic and stable operations \cite{kosut11malicious, liu16masking}. In the literature, the concept of \textit{stealth} FDIA in DC state estimation is first verified in \cite{liu11false} where the attacker can bypass \textit{model-based} Bad Data Detection (BDD) with limited resources, even with protected RTUs. Later, FDIA on AC state estimation is proposed in \cite{hug12vulnerability} along with the vulnerability analysis. FDI attacks for different attack purposes are reviewed in \cite{sayghe20survey} where the corresponding impacts and detection frameworks are partially examined. 

Broadly speaking, three hierarchical approaches can be implemented to counter the cyberphysical attacks, namely protection, detection, and mitigation \cite{ahmed19unsupervised}. The protection is an attack prevention mechanism that can reject common attack attempts. However, it is too hard and costly to protect all the RTUs in the grid and a complete attack rejection is unrealistic \cite{yang14false}. The second stage is attack detection where \textit{model-based} and \textit{data-driven} algorithms are two prevalent methods \cite{musleh19survey}. The third approach is attack mitigation. In \cite{lai2019tri}, a tri-level optimization algorithm is proposed to restore the operation after attack with minimum restoration duration. As the dynamic power system model is hard to track and construct, the model-based detection may easily fail on unseen attack attempts and suffer from detection delay. Consequently, this paper focuses on the detection of cyber attacks using data-driven method while we leave the discussions on the model-based methods in \cite{musleh19survey}.

\subsection{Review on Data-Driven Attack Detection}\label{sec:challenge}

As the number and resolution of the grid measurements are improved, data-driven methods, such as statistical, machine, and deep learning algorithms, are armed to simulate the complex dynamic and uncertainty of the system behavior with few knowledge of the model. Here, we summarize the challenges involved in FDIA detection and review potential solutions of data-driven algorithms in the literature. 

(1) \textbf{\textit{High dimensionality of measurements and features}}:

The dimension of the measured data increases as the topological complexity increases and the anomalies can become hidden and unnoticeable. For example, bus-118 system may have more than 300 features for one sample under DC state estimation while in AC, this number can be boosted to 1k. Due to the static nature of the power system operation and sparse topology, directly changing on the estimated states can only lead to limited measurement variations. Consequently, many statistical and machine learning methods, such as Support Vector Machine (SVM) \cite{xiong2022detection}, Naive Bayesian Classifier (NBC) \cite{cui20flexible}, Decision Trees (DT) \cite{jindal16decision}, and $k$-NNs \cite{zhang06detecting} usually fail to fit in large system with high dimensionality. As a result, dimension reduction or expert feature engineering is usually required for machine learning methodologies in advance. However, the performance of selected features cannot be guaranteed for anomaly detection purpose \cite{pang20deep}. 

(2) \textbf{\textit{Absence of positive measurement samples}}:

Power system operates under the optimal condition in most of the time which leads to the absence of real measured anomalies at training stage. To enrich the balance-labelled dataset in supervised classification, positive measurements are generated artificially \cite{chu20detecting,wu20extreme}. However, in real-time, the attacks can be heterogeneous and there always exists uncertain attack scenarios such that the trained model fails to converge. Due to the rarity of positive samples and the high cost of collecting large number of labeled data, supervised learning is not applicable in practice \cite{pang20deep}. Therefore, generative models, unsupervised, and weakly supervised learnings are proposed. In the generative model \cite{zhang20detecting}, new attack signals can be generated according to the probabilistic distribution of the known attack samples where in unsupervised and semi-supervised learnings, the anomaly detector is mainly trained to describe the normalities and any violations on the acquired knowledge of `normality' can be considered as the outliers. Ref. \cite{ahmed19unsupervised} implements the isolation forest (IF) for unsupervised anomalous measurements detection. Moreover, the normality is examined by semi-supervised mixture Gaussian distribution models in \cite{foroutan17detection} but the anomaly scores are determined by the known attack patterns which weakens its generality. Semi-supervised Support Vector Machine (S3VM) is considered in \cite{ozay16machine} under the assumption that the difference between the number of normal and abnormal samples is not significant. The simulation results of these methods show a higher detection accuracy than common supervised techniques, such as SVM and $k$-NNs, to some extents. However, they usually suffer from high false alarm rate \cite{pang20deep} and roundabout training target \cite{chalapathy19deep}.

(3) \textbf{\textit{Mix of spatial and temporal correlations}}:

The RTU measurements can be considered as a multivariate time series and the anomalies involved can be point-wise, contextual, and cumulative \cite{chandola09anomaly}. In detail, point anomalies are the isolated points that are distinguishable from the majority of the data, e.g. a large but stealthy FDI attack. Using this idea, \cite{moayyed2021image} detects the attacks based on convolutional neural net. Contextual anomalies can only be detected in a certain temporal environment. An extreme example would be the replay attack \cite{mo09secure}. A sub-sequence of the data can be regarded as collective anomalies if it is distinctive to the other instances, e.g., small but stealthy oscillations on the measurements. As the isolated detection on a single measurement mainly focuses on the spatial complexity, it is likely to fail on some attack occasions mentioned above.

Apart from the spatial complexity, it has been reported that the temporal correlations on loads and distributed generators (DG) can reflect on the grid operation and influence on the accuracy of state estimation \cite{zhao18robust}. As a result, prediction-aided detection is considered in \cite{wu20extreme, zhao15forecasting} where the estimated measurement error is found by the predicted states. It should be trivially noticed that, prediction-based detection can benefit on anomaly mitigation by replacing the contaminations with the estimated values. However, as the attacks are usually sparse in the power grid, disregarding all the latest information can also discard the spatial correlations on the normal data, which may deteriorate the detection performance. Another research direction leads to the density-based methodology. Small but successive attacks have been considered in \cite{singh18joint} where data transformation and Kullback–Leibler divergence are implemented to capture the measurement temporal variations while in \cite{yang20bad}, the slopes of the adjacent measurements are classified by decision trees and spectral clustering. However, many of the aforementioned methods only consider the one-step dependence of the measurement dynamics while few can explicitly investigate the spatiotemporal complexity and report the detection accuracy under various unseen attack attempts. 

(4) \textbf{\textit{Robust detection against combined attacks}}:

To facilitate FDIA, the attacker may simultaneously blind a certain part of the measurements by launching an availability or so-called DoS attack, which improves the assault furtiveness and relieves the attack efforts  \cite{pan19cyber,pan16combined, tian20coordinated}. From the resilient point of view, this complicates the detection where the missing data can be either part of the combined integrity-availability attack or caused by the measurement malfunctioning. As a result, static detection algorithm cannot be generalized due to the destruction of the spatial relationships. The goal of the combined attack detection is to give robust estimation on the observed measurements under varying combined attack strengths and missing data ratios. Although the vulnerability analysis has been investigated on the combined attacks, few attention has been paid to the practical detection and mitigation solutions.

\subsection{Deep Autoencoder-Based Anomaly Detection}

In recent years, deep anomaly detection (DNN) has been introduced to numerous practical fields, such as fraud detection and network intrusion \cite{chalapathy19deep}. It also leads to a superior detection performance than the aforementioned traditional methods in power system, such as \cite{he17real,qiu20detection,wu20extreme}. Among all of the DNN applications, deep autoencoder (AE) and its variations have been comprehensively used for feature extraction, dimension reduction, and network pretraining in many machine learning tasks due to its straightforward network structure, unlabelled training requirement, robust, and high-dimensional nonlinear mapping properties \cite{hinton06reducing}. 
As a result, AE-based anomaly detection has become a overwhelming cornerstone. For the attack detection in power systems, point-wise anomaly detection is considered in \cite{wang20detection} while a probabilistic inference is further added by \cite{lin20probabilistic}. Ref. \cite{chen2022data} further extends the AE with Gaussian Mixture Model. The AE detection method explicitly learns the spatial relationships in the measurements while neglecting the temporal correlations. To retrieve the temporal correlations, recurrent neural network (RNN) with different realizations can be considered to replace the dense AE layers. However, the previous applications of deep autoencoder-based detection algorithm assumes the full availability of measurement data, which does not hold in the case of combined integrity and availability attack.   

With the aforementioned challenges (Section \ref{sec:challenge}) in mind, this paper proposes a new Random Denoising LSTM-AE (LSTM-RDAE) based algorithm for power system anomaly detection with the following contributions:
\begin{itemize}
    \item Semi-supervised autoencoder is applied to only learn the normalities of the power system measurement (challenge 2) where the deep network structure can automatically fit on the high-dimensional and nonlinear measurements without prior model knowledge (challenge 1).
    \item We introduce the recurrent LSTM layer in the encoder-decoder network to explicitly extract the temporal correlations among successive measurements (challenge 3). To cope with the availability attacks and improve the robustness, complete random dropout is applied on each sample per epoch (challenge 4). 
    \item We simulate the model by using real-time load profiles on IEEE bus-118. Our method allows a real-time anomaly detection with moderate computational burden. We also consider a broad unseen attack types by considering both the attacker's effort and different attack strength. The simulation results depict that the proposed LSTM-RDAE can outperform the state-of-art semi-/unsupervised machine learning and deep learning methods. 
\end{itemize}

The reminder of the paper is organized as follows. Power system model and combined integrity-availability attacks are formulated in Section \ref{sec:model}. Section \ref{sec:algorithm} introduces the proposed LSTM-RDAE for combined attack detection. The simulation set-up and results are summarised in Section \ref{sec:simulation} while this article ends with a conclusion in Section \ref{sec:conclusion}. 

\section{Problem Formulation} \label{sec:model}

\subsection{State Estimation}

The power grid can be modelled as an undirected graphic network. The generator and demand consist as the nodal buses and the electric power can flow along the transmission lines. A centralized control center can monitor and control the gird operations by collecting nodal and line measurements from RTUs and PMUs. In a large power system, the main objective of the control center is to find the optimal real and reactive powers of each generator such that all the demands are met under the minimal cost while the system constraints are not violated. This generation planning is called as optimal power flow (OPF) \cite{saadat99power}.

State estimation (SE) functions as the core to maintain normal and safe operations in power system. Given redundant measurements, the voltage phasors are estimated through the system equations. Let the system state as $x\in\mathbb{R}^{n}$ and the measurement as $z\in\mathbb{R}^{m}$ with $m>n$, the system equations can be represented as \cite{abur04power}:
\begin{equation}\label{eq:ac-se}
z = h(x)+e\end{equation}
where $e^{T}=\left[e_{1}, e_{2}, \ldots, e_{m}\right]$ is the independent zero-mean measurement errors and define $\mathcal{I} = \{i|i=1,2,\cdots,m\}$ as the measurement index set.

The weighted least square (WLS) algorithm is usually applied to find the states by minimising:
\begin{equation}\label{eq:ac-se-opti}
J(\widehat{x}) = [z-h(\widehat{x})]^{T} E^{-1}[z-h(\widehat{x})]\end{equation}
where $E\in\mathbb{R}^{m\times m }$ is the covariance matrix of the i.i.d. measurement noises and $\widehat{x}$ is the estimated state. As solving \eqref{eq:ac-se-opti} requires iterative minimization, a DC approximation on \eqref{eq:ac-se} is commonly implemented:
\begin{equation}\label{eq:dc-se}
z=H x+e\end{equation}
where $H\in\mathbb{R}^{m\times n}$ is the measurement matrix. In DCSE, the system state only consists of voltage phases: $x=[\theta_2,\cdots,\theta_n]$ where $\theta_1 = 0$ is the reference state. The analytic WLS solution of \eqref{eq:dc-se} is given as
\begin{equation}\label{eq:state_estimation}
    \widehat{x} = ({H}^TE^{-1}{H})^{-1}{H}^TE^{-1}z
\end{equation}

\subsection{Stealthy False Data Injection Attacks}
One of the essential functions of SE is to detect, identify, and eliminate any errors occurring in the measurements. Traditionally, the measurement errors can be caused by sensor inaccuracy and malfunctioning. With the emerging of cyber-physical layers, the SE and BDD are also equipped to detect possible cyber attacks. Firstly, the estimation residual is found as: 
\begin{equation}\label{eq:residual}
    r(z) = \| z-H\widehat{x} \|_2^2
\end{equation}

If \eqref{eq:residual} is normally distributed on different $z$, a Chi-square $\chi^{2}$ algorithm can be applied with $m-n$ freedoms. More directly, a heuristic threshold $\tau$ can be determined according to the normal measurements and the BDD is given as:
\begin{equation}\label{eq: BDD}
D_1(z)=\left\{\begin{array}{ll}
1 & \text { if } r(z) \geq \tau_1 \\
0 & \text { otherwise }
\end{array}\right.\end{equation}
where an attack alarm is indicated by 1.

In this article, an FDIA is considered by directly intruding vector $c\in\mathbb{R}^{n-1}$ on the estimated state $\widehat{x}$ \cite{ahmed19unsupervised}. To successfully achieve the attack goal and also keep stealthy to the BDD, the injected measurement vector should be in the column space of $H$ \cite{liu11false}: 
\begin{equation}\label{eq:perfec_BDD}
    a = Hc,\quad \forall c\in\mathbb{R}^{n-1}
\end{equation}

This stealth FIDA indeed requires a strong condition where the attacker should know the exact system topology and line parameters. In practice, the attacker may also consider to land a series of assaults as long as the attack model
\begin{equation}\label{eq:imperfect_attack}
    \|a-Hc\|_2^2\leq \tau_1-||z-H\widehat{x}||_2^2
\end{equation}
holds \cite{zhao15forecasting,zhao15short}. 

\subsection{Combined Integrity-Availablity Attacks}

Assume a single attack attempt has been imposed on the $i$th bus by amount $\mu$ such that
\begin{equation}\label{eq:state_inject}
c(k)=\left\{\begin{array}{ll}
\mu & \text { if } k=i \\
0 & \text { otherwise }
\end{array}\right.\end{equation}
where $c(k)$ represents the $k$th elements of vector $c$. According to the stealth attack principle \eqref{eq:perfec_BDD}, a possible attack vector can be formulated as:
\begin{equation}\label{eq:perfect_BDD_single}
a = \mu H(:,i)
\end{equation}
where $H(:,i)$ represents the $i$th column of Jacobian $H$. Let $\delta(i) = |H(:,i)|_0$ be the degree of bus $i$, it can be directly shown that the above attack attempt is a \textit{sparsest upper bound} \cite{hug12vulnerability,rahman13false} with respect to the $i$th state. The index set of the contaminated measurements becomes $\mathcal{I}_a = \{j|H(j,i)\neq 0\}$. Generally, the FDIA is intensive and costly as the intruder has to know the system knowledge and impose the attack vector on the real measurement. On the contrary, an availability attack is much cheaper as sensor failure can be easily achieved at both physical and cyber sides. In \cite{pan16combined,pan19cyber,tian20coordinated}, the attacker aims to inject on a particular measurement where the remaining measurements in the \textit{sparsest critical tuple set} are covered to minimize the attack effort. Let $d\in\{0,1\}^m$ be the availability attack vector and $d(k)=1$ indicates that the $k$th measurement is not available, i.e. the availability attack set is $\mathcal{I}_d = \{k|d(k)=1\}$. The measurement matrix will be reduced to $H_r = (I-\text{diag}(d))H$ implying that the $k$th row of $H$ becomes null. In this article, the availability attack strength is defined as $\gamma = |d|_0/m$. Two constraints are considered when applying availability attack on the predefined FDIAs:
\begin{enumerate}
    \item The dropped $kth$ measurement should not be critical, and
    \item $\mathcal{I}_a$ and $\mathcal{I}_d$ are disjoint, i.e. $\mathcal{I}_a \cap \mathcal{I}_d = \phi$.
\end{enumerate}

Condition 1) ensures that the reduced matrix $H_r$ is still observable \cite{abur04power}. Although pseudo-measurements can be imputed on the missing measurements, the deviation of the estimated states can be enlarged if the availability attack continues. It should be pointed out that the control center cannot outcome the desired state contamination $\mu$ on state $i$ \eqref{eq:state_inject} if the measurements in $\mathcal{I}_a$ are unavailable. As shown by condition 2), the low-cost availability attack in this article is introduced to improve the stealthness of the FDIAs and mislead the detection algorithm, which is different to the settings in \cite{pan19cyber,pan16combined}.

To classify different availability attacks, the concept of \textit{missing data} in statistics is adopted. Two measurement blinding mechanisms are considered in particular, named as missing completely at random (MCAR) and missing at random (MAR) \cite{rubin76inference}. In MCAR, all the measurements have the same probability to lose and the missing proportion is a random subset of the measurements $\mathcal{I} \setminus \mathcal{I}_a$ that depends neither on the property nor the value of the measurement. In contrast, MAR or conditional missing is defined by the observed property of individual measurement. For instance, the adversaries may intend to cover certain measurements that are more related to $\mathcal{I}_a$ to keep furtive. Detailed constructions on $\mathcal{I}_d$ will be described in Section \ref{sec:simulation}. The detection algorithm will be trained under MCAR condition whereas both MCAR and MAR will be evaluated during the test stage. Furthermore, we assume that the attacker has full system knowledge including meter measurement data, measurement matrix $H$, and the BDD strategy. Besides, the attacked measurements are assumed to be able to bypass the BDD test \eqref{eq: BDD}.

\section{Detection Methodology}\label{sec:algorithm}

Load profiles are continuously changing in power system due to varying customer behaviours, weather conditions, and dispatch policies. This dependency can be reflected by the correlations of loads at different time which also implies the correlations on states and measurements \cite{wang20physics}. As a result, dynamic behaviour has been considered in \cite{wang20operating, wu20extreme, zhao15forecasting, zhao15short} to aid on anomaly detection by predictive models. However, most of them only consider one-step temporal correlation and the knowledge on the current measurements can be barely applied to the prediction model. In general, the dynamic anomaly detection at time $t$ considering past $T-1$ measurements can be formulated by a look-ahead residual:
\begin{equation}\label{eq:temporal_residual}
    r'(z_t) =\frac{1}{T} \sum_{i=t-T+1}^{t}\| z_i-\widehat{z}_i \|_2^2
\end{equation}
where the dynamic on the measurements is governed by $\widehat{z}_i=l(\widehat{z}_t,\dots,\widehat{z}_{t-T+1})$ for $\forall i=t-T+1,\dots,t$. However, the explicit expression on $l(\cdot)$ is hard to find. As a result, deep learning method is applied to capture the correlations in measurements together with the anomaly detection purpose on measurement $z_t$.

\subsection{LSTM-Autoencoder}

Given a multivariate dataset $\mathcal{Z}=\{z^{<1>},z^{<2>},\dots,z^{<N>}\}$ where $z^{<i>}\in\mathbb{R}^{m}$ has $m$ attributes and a decision space $\mathcal{H}\subset \mathbb{R}^{m_h}$ with $m_h<m$, a commonly-used deep anomaly detection framework is to learn a nonlinear feature mapping $f(\cdot ; \Theta): \mathcal{Z} \mapsto \mathcal{H}$ which can imply anomaly score basing on the learnt knowledge in feature space $\mathcal{H}$. The nonlinear mapping is achieved by stacked network layers, nonlinear activation functions, and the trainable weights and biases ${\Theta} = \{\mathcal{W},\mathcal{B}\}$. 

In a nutshell, an autoencoder is a neural network that is trained to reconstruct its input at the output layer. It consists of two dense networks. For each of the sample $z^{<i>}$, an encoder $h^{<i>} = f_e(z^{<i>};\Theta_e)$ is applied to code the input features into the hidden layer $h^{<i>}\in\mathbb{R}^{m_h}$ with lower dimension while a decoder can then retrieve the inputs by decoding function $\tilde{z}^{<i>} = f_d(h^{<i>};\Theta_d)$ \cite{goodfellow2016deep}. As a result, AE creates a bottleneck for the data where only the significant structural information can go through and be reconstructed. Assumption is made such that the encoded feature can learn nonlinear correlations on the input dataset that are sufficient for separating the anomalies. Since the encoded hidden feature is compact and compressed, only the normal data can be successfully reconstructed by the decoder - thus, anomalies and outliers can be distinguished directly \cite{ruff18deep}:
\begin{subequations}\label{eq:autoencoder}
\begin{equation}
\begin{aligned}
\left\{\Theta_{e}^{*}, \Theta_{d}^{*}\right\} &=\underset{\Theta_{e}, \Theta_{d}}{\arg \min }\left\|z-\tilde{z}\right\|_2^2 \\
&=\underset{\Theta_{e}, \Theta_{d}}{\arg \min } \sum_{{z} \in \mathcal{Z}}\left\|{z}-f_{d}\left(f_{e}\left({z} ; \Theta_{e}\right) ; \Theta_{d}\right)\right\|_{2}^{2}
\end{aligned}
\end{equation}
\begin{equation}
    S(z)=\left\|{z}-f_{d}\left(f_{e}\left({z} ; \Theta_{e}^{*}\right) ; \Theta_{d}^{*}\right)\right\|_{2}^{2}
\end{equation}
\end{subequations}
where ${\Theta_e}$ and ${\Theta_d}$ are training parameters in encoder and decoder networks. Reconstruction error $S(z)$ is the anomaly score assigned to measurement $z$ once the network is trained. Subscript $i$ is omitted in \eqref{eq:autoencoder} for brevity. 
Similarly to the BDD \eqref{eq: BDD}, the decision threshold $\tau_2$ can be found according to the distribution of the reconstruction error on the validation set and the detection principle is set as:
\begin{equation}\label{eq: autoencoder_detection}
D_2(z)=\left\{\begin{array}{ll}
1 & \text { if } S(z) \geq \tau_2 \\
0 & \text { otherwise }
\end{array}\right.\end{equation}

While the autoencoder has proven to be useful for anomaly detection where only the normal measurements are collected to train the network, it can only detect point outliers and overlook the temporal correlations between samples, e.g. the contextual and cumulative anomalies. To explicitly extract the temporal correlations, recurrent neural network (RNN) is used to replace the feedforward networks in both encoder and decoder. In RNN, the impact of the previous input can be recorded and new features are learnt recursively in each training sample \cite{goodfellow2016deep}. Perhaps, the most effective recurrent model in practice is LSTM \cite{hochreiter97long}. To deal with the long-term gradient vanishing and explosion problem, LSTM can automatically update and forget the state in each cell.

\begin{figure}[ht]
     \centering
     \begin{subfigure}[b]{0.48\textwidth}
         \centering
         \includegraphics[width=\linewidth]{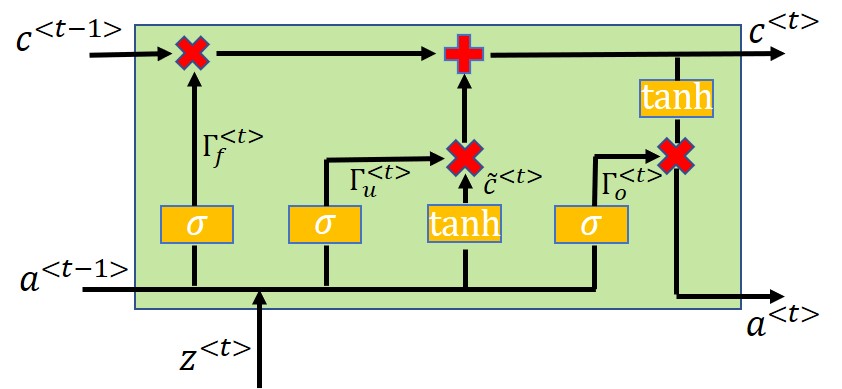}
         \caption{}
         \label{fig:y equals x}
     \end{subfigure}
     \hfill
     \begin{subfigure}[b]{0.48\textwidth}
         \centering
         \includegraphics[width=\linewidth]{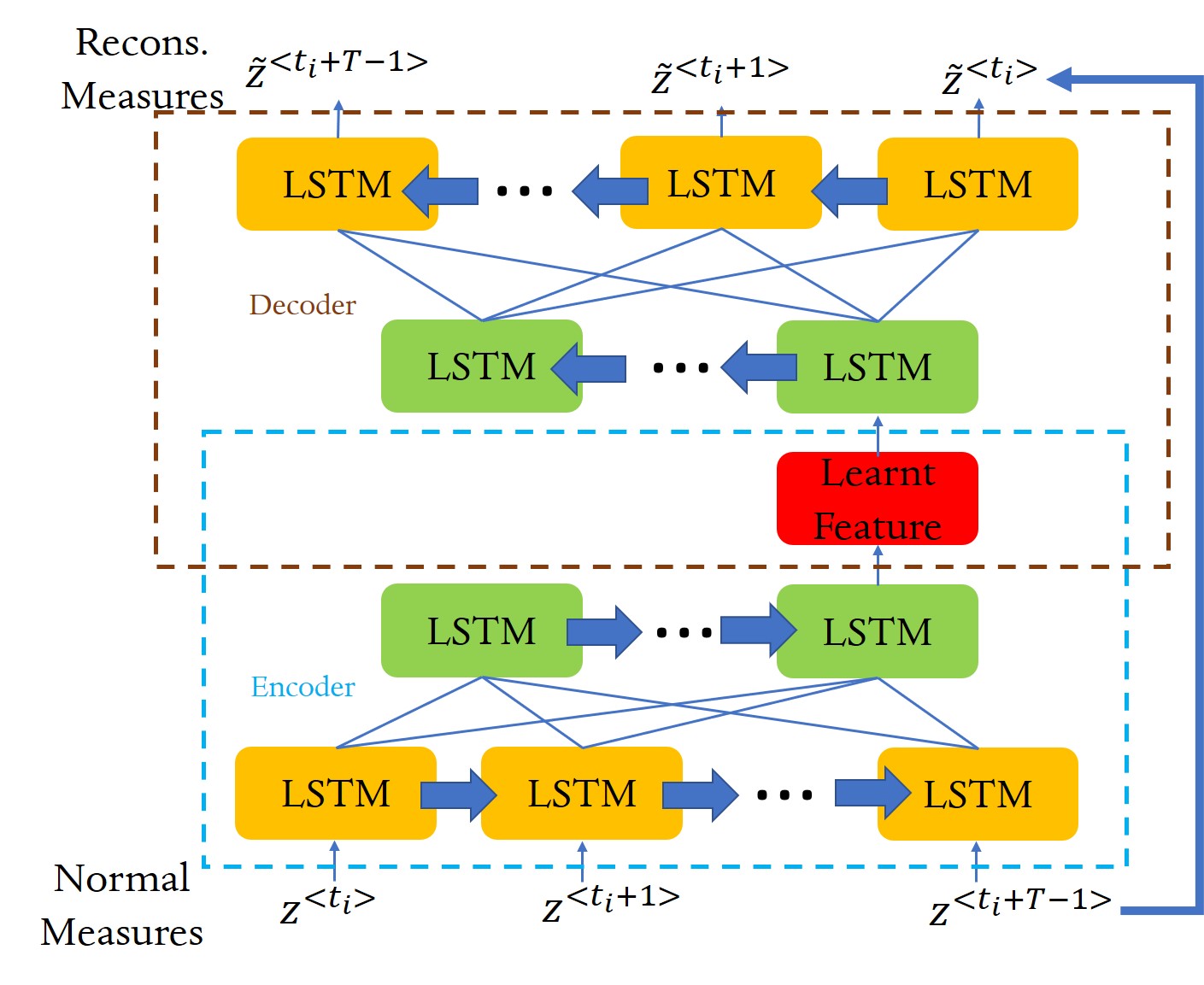}
         \caption{}
         \label{fig:three sin x}
     \end{subfigure}
        \caption{LSTM structure: (a).a single LSTM cell, and (b).LSTM-AE}
        \label{fig:LSTM_structure}
\end{figure}

Consider a length $T$ continuous subset of $\mathcal{Z}$: $\mathcal{Z}_i = \{z^{<t_i>},$ $z^{<t_i+1>},\cdots, z^{<t_i+T-1>} \}$ with $t_i\leq N-T+1$. An LSTM cell for input $z^{<t>}$ can be calculated as Fig.\ref{fig:LSTM_structure}(a) and \eqref{eq:lstm}. In \eqref{eq:lstm}, $W_*$ are the kernels, $b_*$ represents the biases, $\sigma$ and $\text{tanh}$ are commonly used as the sigmoid and hyperbolic tangent activation functions. In detail, at time $t$, state $c^{<t>}$ and the activation $a^{<t>}$ are updated by the current input $z^{<t>}$, the previous state $c^{<t-1>}$, and previous activation $a^{<t-1>}$. The LSTM cells are initiated statelessly by $a^{<t_i-1>}=c^{<t_i-1>} = 0$. Following \eqref{eq:lstm}, the useful information in the past can always be kept by the update gate and the unworthy one will be forgotten by the forget gate. In our problem, this suggests that the temporal correlations that are critical for normality representation are extracted through a window of length $T$.
\begin{equation}\label{eq:lstm}
\begin{array}{c}
\text { Candidate state: } \tilde{c}^{<t>}=\tanh \left(W_{c}\left[a^{<t-1>}, z^{<t>}\right]+b_{c}\right) \\
\text { Update gate: } \Gamma_{u}^{<t>}=\sigma\left(W_{u}\left[a^{<t-1>}, z^{<t>}\right]+b_{u}\right) \\
\text { Forget gate: } \Gamma_{f}^{<t>}=\sigma\left(W_{f}\left[a^{<t-1>}, z^{<t>}\right]+b_{f}\right) \\
\text { Output gate: } \Gamma_{o}^{<t>}=\sigma\left(W_{o}\left[a^{<t-1>}, z^{<t>}\right]+b_{o}\right) \\
\text { State update: } c^{<t>}=\Gamma_{u}^{<t>} \tilde{c}^{<t>}+\Gamma_{f}^{<t>} c^{<t-1>} \\
\text { Output: } a^{<t>}=\Gamma_{o}^{<t>} \tanh \left(c^{<t>}\right)
\end{array}\end{equation}

Fig.\ref{fig:LSTM_structure}(b) illustrates the structure of the LSTM autoencoder (LSTM-AE) where each layer represents an unfolding graph of one LSTM cell for sample $\mathcal{Z}_i$ and it is further stacked to have a deep structure. The LSTM-AE can be trained similarly to \eqref{eq:autoencoder} where the anomaly score for $\mathcal{Z}_i$ is the mean over sample length $T$:
\begin{equation}\label{eq:lstm_score}
    S(\mathcal{Z}_i)=\frac{1}{mT}\sum_{j=0}^{T-1}\left\|{z}^{<t_i+j>}-l_{d}\left(l_{e}\left({z}^{<t_i+j>} ; \Theta_{e}^{*}\right) ; \Theta_{d}^{*}\right)\right\|_{2}^{2}
\end{equation}
where $l_e$ and $l_d$ represent the LSTM encoder and decoder mappings. Consequently, \eqref{eq:lstm_score} gives a feasible representation on \eqref{eq:temporal_residual} where the measurement dynamics are embedded in $l(\ \cdot \ ; \Theta)$ during training. 

\subsection{Random Dropout Layer}

Conventionally, denoising autoencoder (DAE) is used to train the network to learn more `intelligently' on the hidden features instead of trivial mapping. Noisy layer, such as Gaussian noise mask and dropout mask \cite{vincent10stacked}, is added directly on the input and a nonlinear mapping is trained to reconstruct the clean input signal:
\begin{equation}
    f_{DAE}: f_{d}\left(f_{e}\left({\overline{z}} ; \Theta_{e}^{*}\right) ; \Theta_{d}^{*}\right) \mapsto z
\end{equation}
where $\overline{z}$ is the corrupted version of $z$. In this article, the dropout mask is applied to quantify the process of missing mechanism on normal data. Though in real life, the missing ratio can be known by the control center whenever the current measurements are available, its value $\gamma$ can be varying continuously. Unavoidably, this requires the control center to have the detection model on each missing ratio case which is computational unattractive, lack of redundancy, and may also cause overfitting problem. Inspired by DAE \cite{vincent08extracting}, the basic idea of the completely random dropout on the input layer is to use the Monte-Carlo method to model all possible data missing ratios in range $[d_{min},d_{max}]$. 

During each training epoch, a random dropout mask $D_i$ is applied on the $ith$ sample to give the corrupted counterpart:
\begin{equation}
    \overline{\mathcal{Z}}_i = \mathcal{Z}_i - \mathcal{Z}_i \circ D_i
\end{equation}
where $\circ$ is the Hadamard product and $D_i\in\mathbb{R}^{m\times T}$ with $D_i(j,k) = 1$ indicating that the $j$th attributes in the $k$th measurements is unavailable during the current loop. The missing ratio and attribution in $D_i(:,k)$ is random and independent on different $i$ and $k$ to simulate the MCAR condition. Consequently, after sufficient trainings (epochs), the LSTM-RDAE can fit on normal measured sequence under various missing ratios. 

\begin{figure}[ht]
  \begin{center}
  \includegraphics[width=0.5\linewidth]{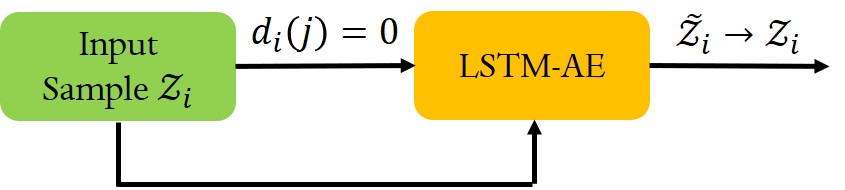}
  \caption{LSTM-RDAE configured for single temporal sample.}
  \label{fig:LSTM_DAE_structure}  
  \end{center}
\end{figure}

To control the complexity in simulation, we assume that the previous missing measurements have been imputed or estimated for further operations such as dispatch organization and load forcasting. As a result, only the incoming measurement is detected with missing attributes in real time: 
\begin{equation}
    \overline{z}^{<t_i+T-1>} = {z}^{<t_i+T-1>} \circ \left(\textbf{1} - d_i \right) 
\end{equation}
where $d_i \in \{0,1\}^{m}$ by controlling the missing strength $d_{min}$ $\leq |(d_i(j)|_0/m \leq d_{max}$. Fig.\ref{fig:LSTM_DAE_structure} illustrates the connection between random input dropout to the conventional LSTM-AE.

\begin{remark}
We refer the autoencoder-based anomaly detection as semi-supervised as only the normal measurements are required to train the network. This has fundamental differences to the unsupervised learning as the exact positive and negative labels cannot be known in advance \cite{pang20deep}. 
\end{remark}

\begin{remark}
Though sharing with certain similar ideas, our method, as a new DAE \cite{vincent08extracting} application on modelling normal missing data, is different to \cite{gal16theoretically} where the fixed forward and recurrent dropout ratios are used to approximate the variational inference. It also varies from the sparse LSTM-AE \cite{kieu19outlier} and ensemble RandNet \cite{chen17outlier} in which the multiple networks with random and independent disconnections in the hidden layers are applied.
\end{remark}

\subsection{Discussions on LSTM-RDAE} 

\begin{figure}[ht]
  \begin{center}
  \includegraphics[width=0.5\linewidth]{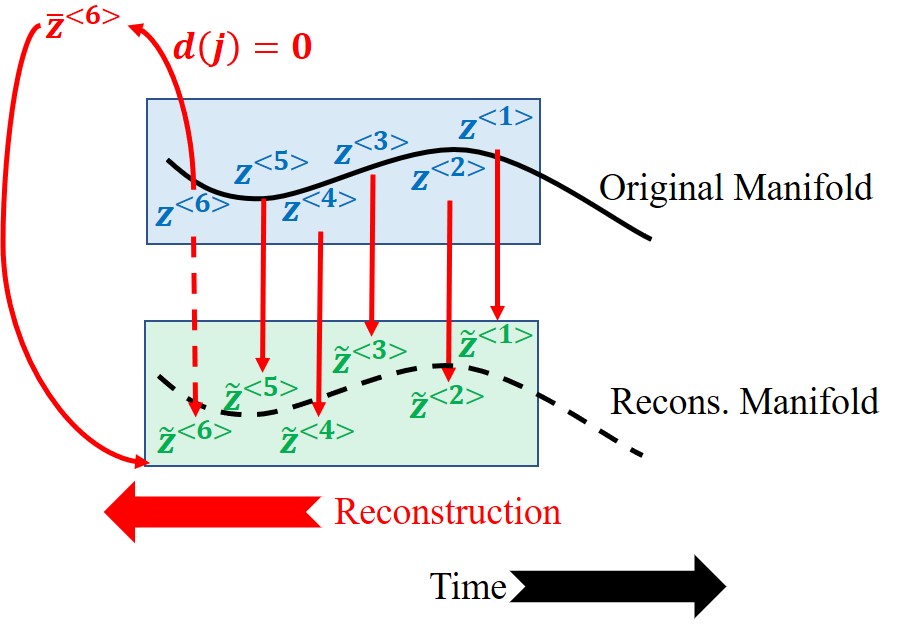}
  \caption{LSTM-RDAE formulated as a manifold reconstruction problem.}
  \label{fig:LSTM_DAE_manifold}  
  \end{center}
\end{figure}

In this section, we intuitively explain why applying sequential information can improve the performance of DAE for combined attack detection. Referring to the manifold assumption \cite{vincent10stacked}, high-dimensional data locates around some certain low-dimensional manifolds (such as the solid black curve in Fig.\ref{fig:LSTM_DAE_manifold}). The autoencoder can be interpreted as regenerating a similar manifold (dashed black curve) to catch the projected sequential data. The encoder representation $l_{e}(\ \cdot \ ;\Theta_e)$ can thus be argued as a certain projection pattern at this lower manifold space. When dropout is applied, the deviation between $\overline{\mathcal{Z}}_i$ and $\mathcal{Z}_i$ is amplified in the lower space, i.e. the distance to the manifold increases. As shown by Fig. \ref{fig:LSTM_DAE_manifold}, the previous uncorrupted measurements can help guide the reconstructed manifold up to the incoming corruption by learning their temporal correlations even under various dropout ratios.     


\section{Experiment Set-ups and Results} \label{sec:simulation}

In this section, we evaluate the performance of LSTM-RDAE algorithm for combined attack detection problem.

\subsection{Data Acquisition, Refinement, and Analysis}
We use some common settings in literature to prepare the training dataset. To simulate the most realistic situation and obtain high-resolution real-time measurements and states, 11 distinct European regional load profiles in year 2015-2017 with 15min observation interval are collected from an open-source dataset \cite{hirth19time}. Firstly, the missing data is roughly imputed by the last-day's load at the same time. To have a reasonable profile range, for each regional load pattern, a random maximum load consumption $l_{i, \max}$ is applied \cite{zhang20topology}:
\begin{equation}
L^{<k>}_i=l_{i,\max} \frac{l^{<k>}_i}{\max_{k} \left(l_i\right)}, \quad i = 1,2,\dots 11
\end{equation}
where $l^{<k>}_i$ is the load profile in region $i$ at time $k$, $l_{i,\max } \in [0.25,2.75]$ is the random maximum power for region $i$ in $p.u.$. $\max_{k} \left(l_i\right)$ is the maximum power at region $i$, and $L^{<k>}_i$ is the rescaled regional load profile.

\begin{figure}[ht]
  \begin{center}
  \includegraphics[width=0.5\linewidth]{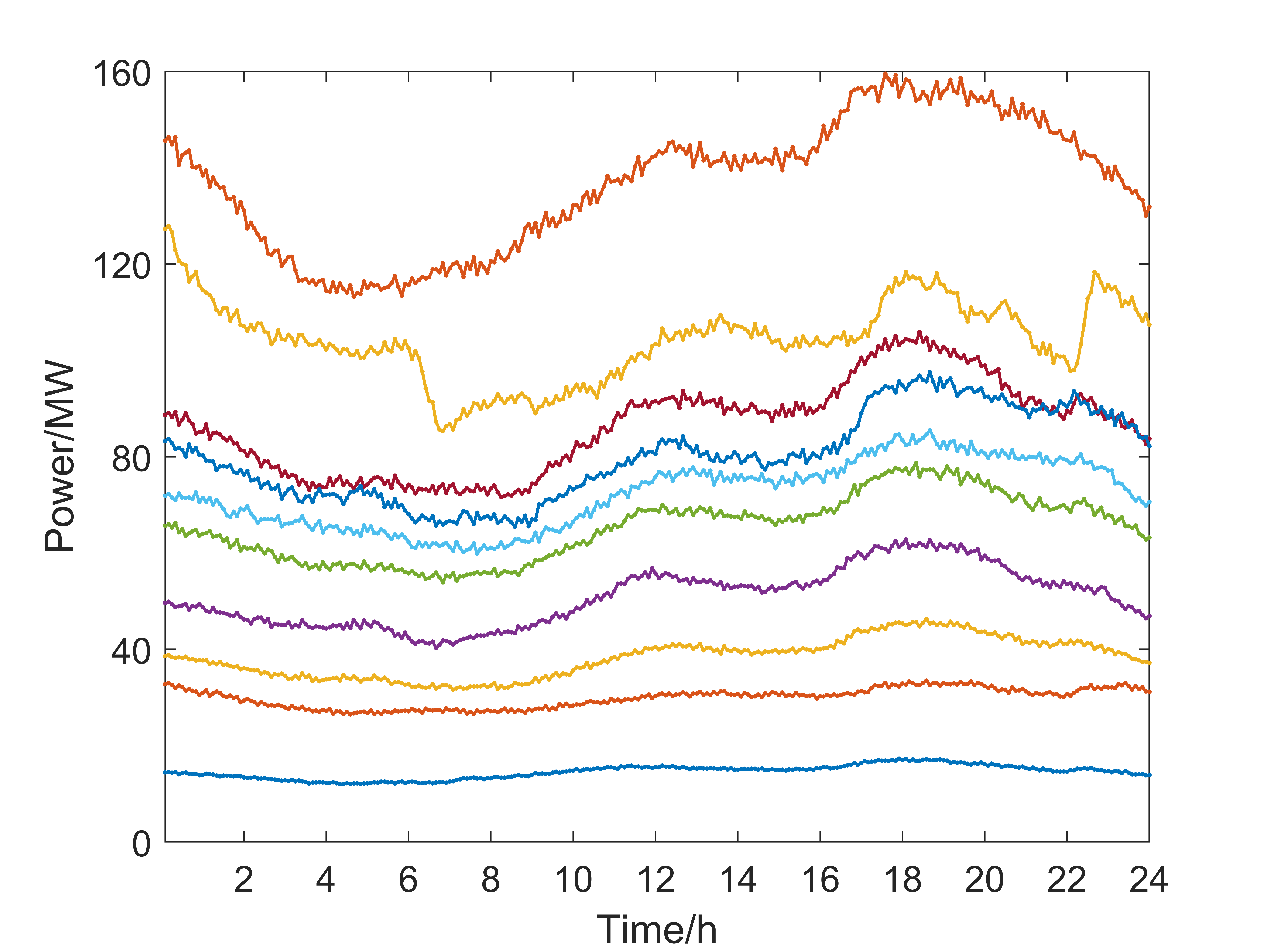}
  \caption{One day load profile for the first 10 out of 99 loads in bus-118 system}
  \label{fig:load_profile}  
  \end{center}
\end{figure}

IEEE bus-118 test system \cite{zimmerman11matpower} is used as the simulation platform which contains 99 loads and 54 generators in total. To map the 11 distinct load profiles into 99 load buses and keep their individual pattern, symmetric Dirichlet distribution \cite{wang20detection} with a low distribution parameter $a=0.2$ is applied. Furthermore, as in general the state estimation can be processed in less than 5min, two more load points are interpolated within the successive 15min data with 2\% variations added. Consequently, each bus has 288 load data a day accounting for 210528 temporally continuous load data in total, e.g. in Fig.\ref{fig:load_profile}. After obtaining the $\mathcal{L}\in\mathbb{R}^{99\times 210528}$ loads, AC-OPF is operated with the specified active power demand. The power factor is found as the default value with 5\% variations \cite{singh18joint, zhang20topology}. To increase the randomness, stepped quadratic generator costs are applied at each node. The remaining parameters, system topology, and operational constraints are set as defaults as in MATPOWER bus-118 description file \cite{zimmerman11matpower}. At each time $k$, the AC-OPF gives the solution to line power flows $P_f\in\mathbb{R}^{186}$ and power generations $P_g\in\mathbb{R}^{54}$. Then the modified measurement vector can be further found by
\begin{equation}
    z = \left[\begin{array}{c}
        C_{g} P_{g}-P_{d}  \\
        P_f 
    \end{array}\right]
\end{equation}
where $C_{g}\in\mathbb{R}^{118\times54}$ is the generator incidence matrix and $C_{g}{(i,j)} = 1$ if the $jth$ generator locates at bus $i$, otherwise $C_{g}{(i,j)} = 0$. $P_d \in \mathbb{R}^{118}$ is the demand vector. The augmented measurement $z=\left[P_{I} ; P_{F}\right] \in\mathbb{R}^{304\times 210528}$ with $1\%$ Gaussian measurement noise added will be used as the input to the neural network. Moreover, the \textit{min-max scaling} is applied on the raw measurement data into range $[0,1]$ across samples.

As discussed in Section \ref{sec:model}, different buses have different venerability degrees $\delta$ under FDIAs which can also reflect on the attack efforts. For the IEEE bus-118 system under DC assumption, the degree can vary from $\delta = 3$ at bus 10, 73, 87, 111, 112, 116, and 117 up to $\delta = 22$ at bus 49. If $\delta = 3$, the targeted bus is connected to one other bus and changing the state will only effect on the two buses’ states and the line power between them which makes the attack hard to detect in general. 


\subsection{Model Settings and Parameter Tuning}

In general, it is not conspicuous to train the semi supervised LSTM-RDAE algorithm before the detection accuracy is counted by applying the scoring credit \eqref{eq:lstm_score} since the anomalies are not supposed to know in advance. Thus, as one of the baseline methods, we pretrain an vanilla autoencoder method with full measurement information. Three metrics are considered throughout the training and evaluation procedures: (a). True Positive Rate (TPR) which counts for the ratio of the predicted positive samples over the actual positive samples, (b). False Positive Rate (FPR) which counts for the predicted positive samples over the actual negative samples (the false alarm), and (c). the $F_1$ score which is defined by:

(1). The precision $Pre$: the proportion of the correctly predicted anomalies in all the predicted anomalies
\begin{equation}
\label{precision} 
Pre = TP / (TP + FP)
\end{equation}

(2). The recall $Rec$: the proportion of the correctly predicted anomalies in all the true anomalies
\begin{equation}
\label{specificity} 
Rec = TP / (TP + FN)
\end{equation}

(3). The $F1$-score: the harmonic mean of $Pre$ and $Rec$
\begin{equation}
\label{F1-score} 
F_1\text{ score} = 2 \times Pre \times Rec/(Pre + Rec)
\end{equation}

Note that for a competitive classifier, the $F_1$-score should approach to 1. The core hyperparameters in AE network is the hidden feature dimension, the number of layers, and the activation functions. We test a wide range of these parameters where nonlinear activations and compression hidden features are designed in particular to avoid trivial feature extraction. We then give some benefits to vanilla AE by supervised training. In detail, the hyperparameters are tuned using grid search method by evaluating its best $F_1$ score. The optimal hyperparameters are then implemented directly to the LSTM-RDAE. Table \ref{table:para} records the hyperparameters for LSTM-RDAE guided by vanilla AE where the training epoch is tuned to have similar loss as AE. The sample length $T = 6$ is chosen since the successive attack is usually assumed to last for at most three steps \cite{yang20bad}. Further enlarging the sample length will dilute the impact of isolated anomalous point thus reduce the detection accuracy. Accordingly, the training samples in LSTM-RDAE are separated by sliding window \cite{yang20lstm} of length-6, step-1 so that a $\mathcal{Z}_{LSTM}\in \mathbb{R}^{304 \times 6 \times 210523}$ tersor can be constructed.

\begin{table}[ht]
\centering
\scalebox{1}{
\begin{tabular}{lc}
\hline
\multicolumn{1}{c}{\textbf{Parameter}} & \textbf{Value}               \\ \hline
\textbf{Layer structure}  & 304-512-256-256-512-304                   \\
\textbf{Batch size}                    & 400                      \\
\textbf{Sample length}                 & 6
\\
\textbf{Number of epochs}              & 1500                         \\
\textbf{Dropout rate (input)}          & $[0,0.2]$    \\
\textbf{Dropout rate (hidden)}         & 0.005                         \\
\textbf{Optimizer}                     & Adam                         \\
\textbf{Normalisation}                 & min-max normalization                        \\ 
\textbf{Learning rate}                 & 0.0001                        \\ \hline
\end{tabular}
}
\caption{Hyper-parameters of the Proposed LSTM-RDAE}
\label{table:para}
\end{table}

The random input dropout rate is set as $d_{min}=0$ to simulate the condition where no availability attack is imposed. In a bus-14 example, 3 up to 12 out of the 55 total measurements can be dropped to hide an FDIA on a single measurement \cite{pan19cyber} and this number can rise to 36 if a coordinated attack is considered \cite{tian20coordinated}. Without losing generality, the maximum missing ratio is set as $d_{max}=0.2$ in our situation. Moreover, feedforward dropouts \cite{gal16theoretically} are added in the encoder hidden layers to avoid overfitting in LSTM-RDAE. The training and validation process is operated on the first-year's data which is randomly partitioned into $\mathcal{Z}_{LSTM}^{Train}:\mathcal{Z}_{LSTM}^{Valid} = 0.8:0.2$ and the trained network will be evaluated solely on the second year's data for both normality and anomaly. To sum up, attacked measurements of different kinds are generated only on $\mathcal{Z}_{LSTM}^{Test}$.


\subsection{Stealthy Combined  Integrity-Availability Attacks With Limited Resources} \label{sec:detection_strategy}

In this section, we follow the similar attack patterns in \cite{ahmed19unsupervised} where only a specific single estimated state on every bus except bus 69 (as the slack) can be contaminated according to stealth attack model \eqref{eq:perfect_BDD_single}, as we are aiming to find the most general detection algorithm. Each single state can vary in set $\pm\left[3\%,5\%,7\%,10\%,15\%,20\%,30\%\right]$. For the one-shot (point) attacks, the averaged metric is calculated over all the 117 attack cases under different FDIA strengths and availability attack ratios.

\subsubsection{Detection Strategy and Model Evaluations}

\begin{figure}[ht]
  \begin{center}
  \includegraphics[width=0.5\linewidth]{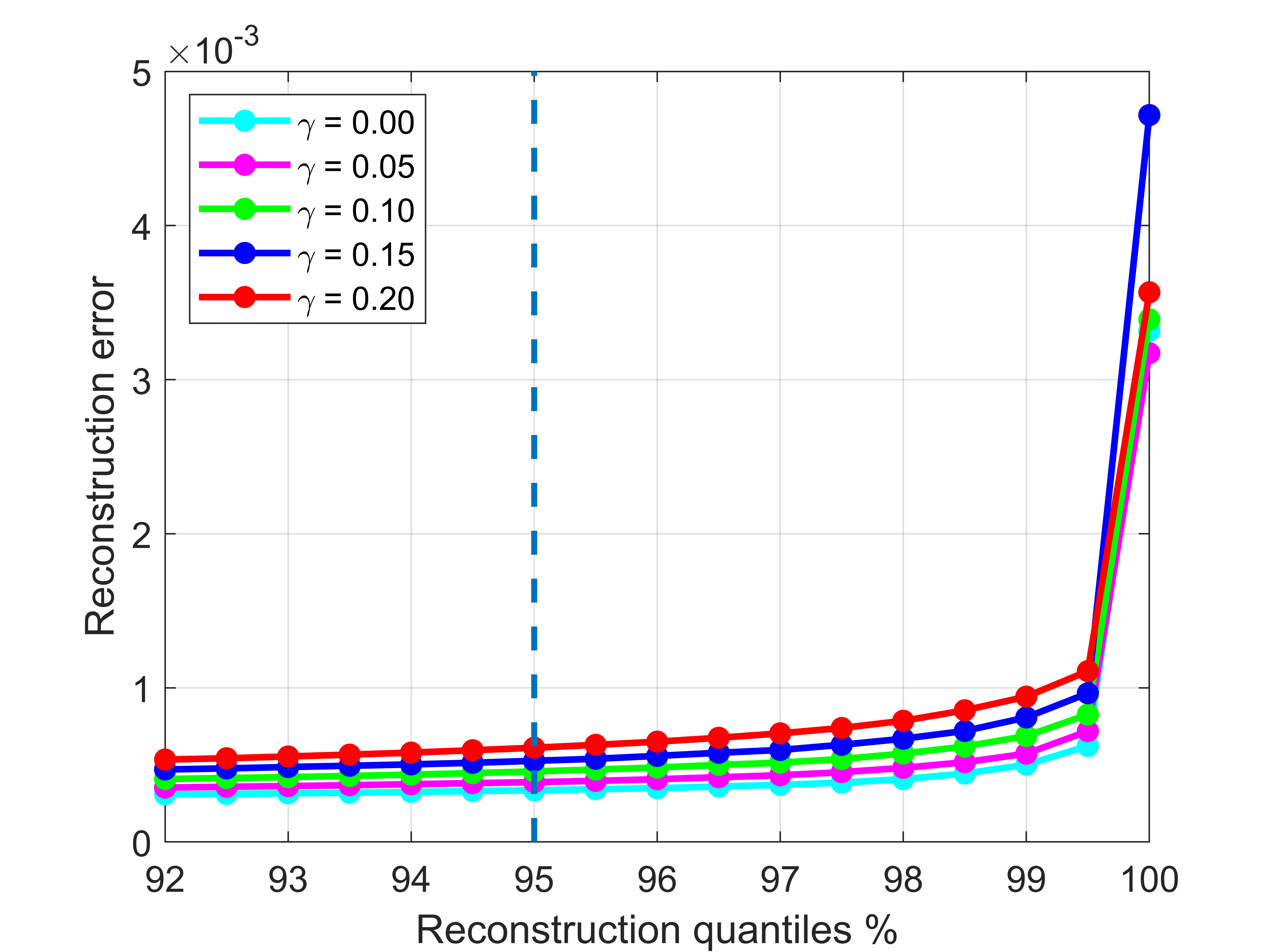}
  \caption{Reconstruction error on the normal measurements in $\mathcal{T}^{valid}_{LSTM}$ with different missing ratios $\gamma$.}
  \label{fig:valid_residual}  
  \end{center}
\end{figure}

\begin{figure}[ht]
  \begin{center}
  \includegraphics[width=0.5\linewidth]{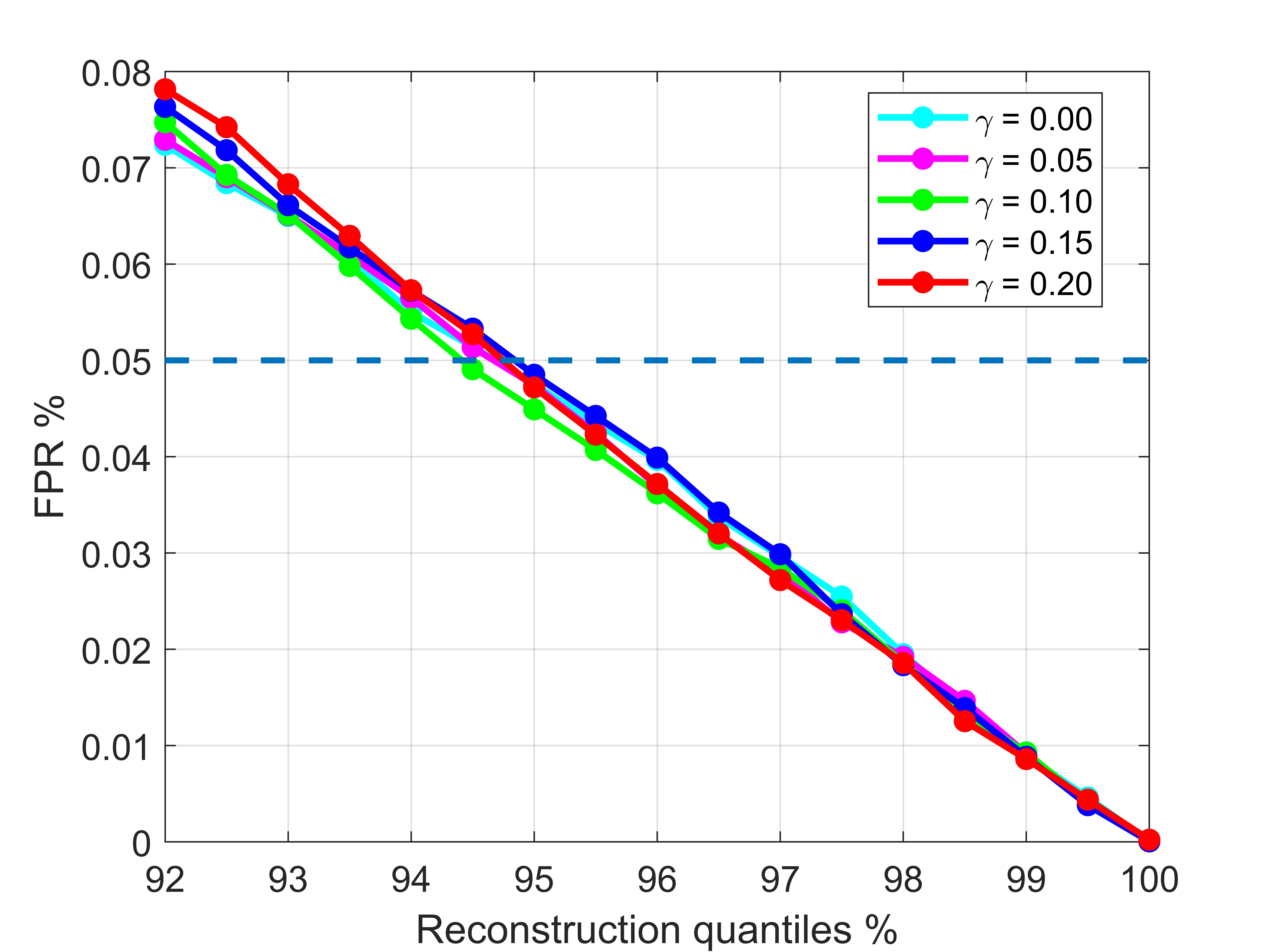}
  \caption{$FPR$s on the normal measurements in $\mathcal{T}^{test}_{LSTM}$ with different missing ratios $\gamma$.}
  \label{fig:false_positive_rate}  
  \end{center}
\end{figure}

After training the LSTM-RDAE network, the reconstruction error \eqref{eq:lstm_score} on the normal validation set $\mathcal{T}^{Valid}_{LSTM}$ are found under different availability attack ratios $\gamma$ and sorted in an ascending order. Basing on the error distributions, their $\alpha$th quantile can be found individually. As shown by Fig.\ref{fig:valid_residual}, except the large deviations at $\alpha=100\%$, the reconstruction errors are similar under different $\gamma$. Besides, the lowest anomaly score is achieved when there is no missing measurements ($\gamma = 0.00$). Fig. \ref{fig:false_positive_rate} illustrates the $FPR$ on normality $\mathcal{T}_{LSTM}^{test}$ according to the predefined thresholds in Fig.\ref{fig:valid_residual} where stable false alarms can be observed under different availability attacks. We can also observe that the $\alpha$th $FPR$ is equivalent to their corresponding $\alpha$th quantile, i.e. $FPR(\alpha)+\alpha \approx 1$, regardless of the varying ratios which suggests that the proposed LSTM-RDAE can overcome the overfitting problem and give a better representation on the unseen measurements. 

In real-time anomaly detection, $\gamma$ is likely to vary continuously that is different from the five cases considered in Fig. \ref{fig:valid_residual} and \ref{fig:false_positive_rate}. To further test the robustness, five available attack ranges are defined accordingly by Table \ref{table:missing_ratio}. For example, if the current measurement has 20 unavailable attributes, then the problem will be solved by the principle defined at $\gamma = 0.05$.  
\begin{table}[ht]
\centering
\scalebox{1}{
\begin{tabular}{cc}
\hline
\multicolumn{1}{c}{\textbf{Availability Attack Range}} & \textbf{Detection Standards}               \\ \hline
$[0,0.025)$ &  $\gamma = 0.00$                   \\
$[0.025,0.075)$                    & $\gamma = 0.05$                      \\
$[0.075,0.125)$                    & $\gamma = 0.10$\\
$[0.125,0.175)$                    & $\gamma = 0.15$\\
$[0.175,\infty)$                   & $\gamma = 0.20$\\ \hline
\end{tabular}
}
\caption{Ranges and Detection Standards for Different Missing Ratios}
\label{table:missing_ratio}
\end{table}

\begin{figure}[ht]
  \begin{center}
  \includegraphics[width=0.5\linewidth]{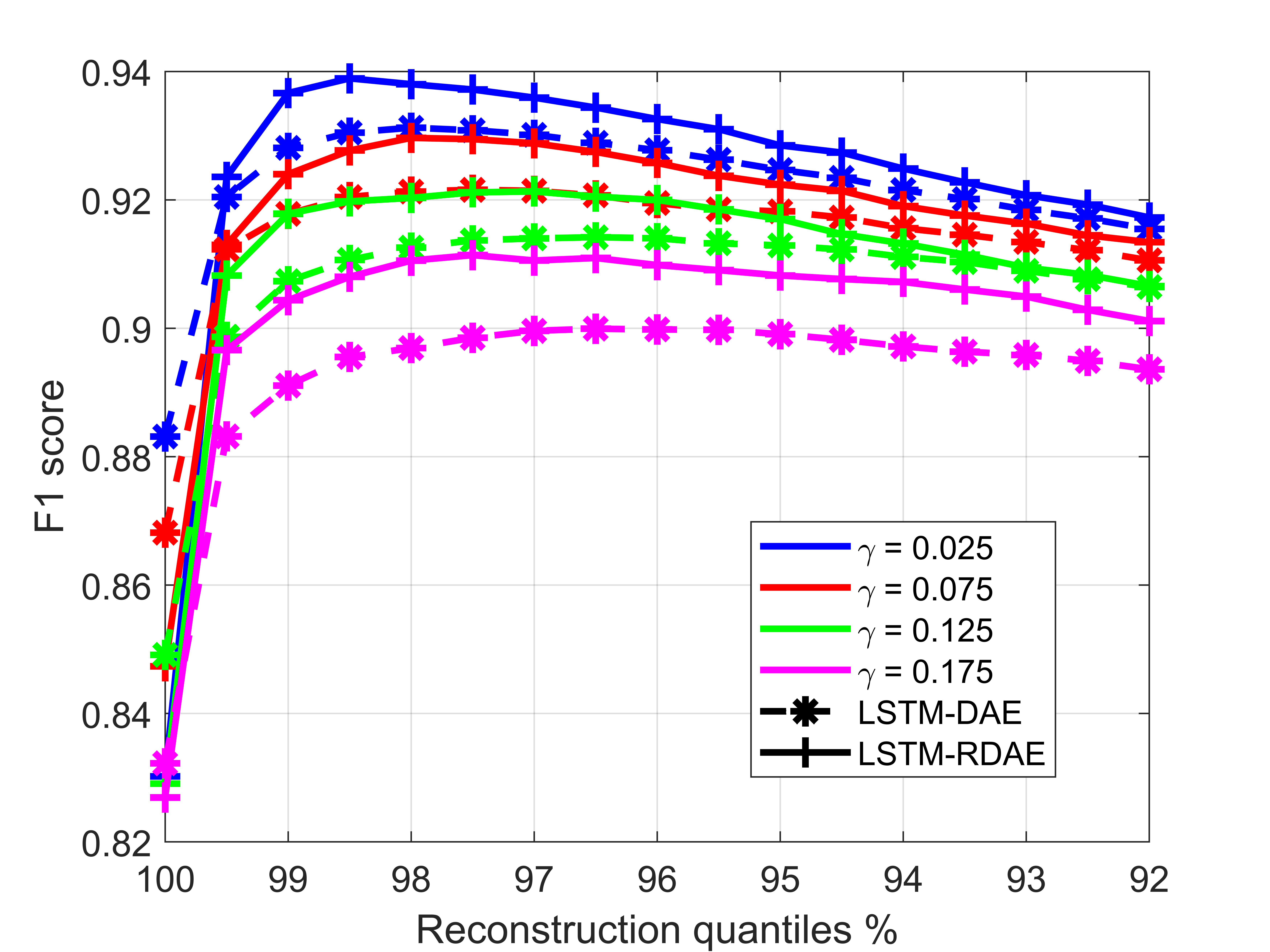}
  \caption{Sensitivity of quantiles: averaged $F_1$ scores under 10\% FDIAs and different availability attack ratios $\gamma$. The LSTM-DAE is trained at $\gamma$ whose detection threshold is found according to Table \ref{table:missing_ratio}.}
  \label{fig:compare_RDAE_DAE} 
  \end{center}
\end{figure}

\begin{figure}[ht]
  \begin{center}
  \includegraphics[width=0.5\linewidth]{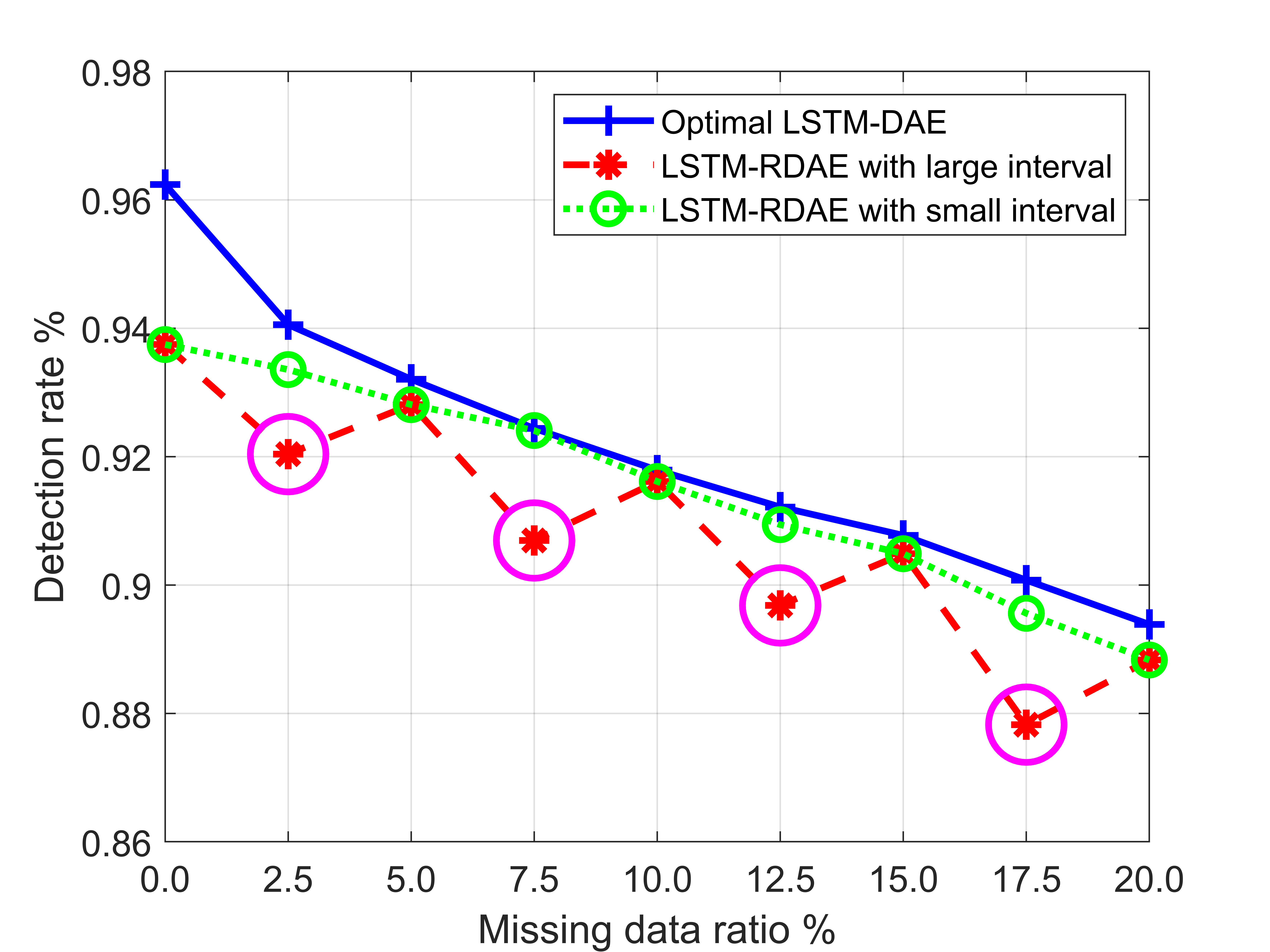}
  \caption{Sensitivity of availability attack detection range: averaged detection rates under 10\% FDIAs and different availability attack ratios $\gamma$. Both the model and detection threshold of LSTM-DAE are found at $\gamma$.}
  \label{fig:compare_quantile_range} 
  \end{center}
\end{figure}

The differences between the LSTM-AE with and without the proposed random dropout layer are compared in Fig. \ref{fig:compare_RDAE_DAE} and \ref{fig:compare_quantile_range}. In detail, Fig.\ref{fig:compare_RDAE_DAE} compares the averaged $F_1$ scores with $10\%$ attack on each bus. The LSTM-DAEs are trained on the exact five $\gamma$ values in Table \ref{table:missing_ratio} and both DAEs and RDAE are to follow the detection scheme in Table \ref{table:missing_ratio}. To test the robustness, the two detection algorithms are tested at the boundaries of the available attack ranges, which stands for the hardest possible detection cases, i.e. at $\gamma \in \Gamma_b = [0.025,0.075,0.125,0.175]$. In general, the detection performance is deteriorated when $\gamma$ increases. This observation obeys the assumption that imposing availability attack on the FDIA can improve the assault stealthy level. Furthermore, the $F_1$ score with random dropout layer (LSTM-RDAE) is higher than it with the exact missing ratio preference (LSTM-DAE) in all cases, inferring that the proposed method is more accurate and robust.

Referring to Fig.\ref{fig:false_positive_rate} and \ref{fig:compare_RDAE_DAE}, the quantile $\alpha$ is another parameter influencing the detection metrics. Decreasing the quantile can improve the $TPR$ in the cost of misclassifying more normal measurements. As suggested by \cite{musleh19survey, wang20detection}, $FPR=5\%$ is a common choice to balance the trade-off of the $F_1$ score. In the following discussions, all the comparisons are operated by controlling $5\%$ false alarms on each method unless specified. 

Fig.\ref{fig:compare_quantile_range} investigates the effect of different interval lengths of the available attack ratio ranges in Table \ref{table:missing_ratio}. The optimal LSTM-DAE is simulated where each network is trained and the detection strategy is determined particularly by their individual available attack ratio. As a result, it can be considered to have the best detection performance. As in Fig.\ref{fig:compare_RDAE_DAE}, the red dotted curve in Fig.\ref{fig:compare_quantile_range} follows the detection strategy in Table \ref{table:missing_ratio} by LSTM-RDAE. The pink cycles highlight that the worst detection rates occurring at the boundaries. If the detection principles are further investigated explicitly at $\gamma\in\Gamma_b$, the detection rates can be significantly improved (green curve in Fig. \ref{fig:compare_quantile_range}). In general, we can always achieve a comparable detection performance using LSTM-RDAE by retrieving the exact detection principle, except when $\gamma = 0$. Unlike LSTM-DAE, finding a new detection standard on a certain $\gamma$ does not require to retrain the network and the computational time is less than 0.5s in average. 


\subsubsection{Model Verification}

In this section, we evaluate the performance of the LSTM-RDAE compared with other state-of-art deep and machine learning algorithms. In particular, we investigate how the temporal information can support the detection process with and without availability attacks. As the detection on combined attacks are barely researched in literature, we modify the existing method on FDIA detection as follows:
\begin{itemize}
    \item Baseline 1: DAE by adding denoising layers in the AE-based detector in \cite{wang20detection} and \cite{chen2022data};
    \item Baseline 2: LSTM-AE by replacing the dense layers with the LSTM layers in \cite{wang20detection};
    \item Baseline 3: One-Class Support Vector Machine (OC-SVM) as an extension to the S3VM in \cite{ozay16machine};
    \item Baseline 4: Isolation Forest (IF) in  \cite{ahmed19unsupervised}.
\end{itemize}

\begin{table*}[ht]
\centering
\scalebox{1}{
\begin{tabular}{l|ccccccc}
\hline
\multicolumn{1}{c|}{\textbf{Algorithms}} & $\%\mu = 0.03$ & $0.05$ & $0.07$ &$0.10$ & $0.15$ & $0.20$ & $0.30$            \\ \hline
\textbf{DAE}          & \textcolor{brown}{0.8011} & 0.8629 & 0.9108 & 0.9306 & 0.9506 & 0.9602 & 0.9726 \\
\textbf{LSTM-AE}      & \textcolor{red}{0.8334} & 0.9019 & 0.9270 & 0.9463 & 0.9627 & 0.9718 & 0.9788 \\
\textbf{LSTM-RDAE}    & \textcolor{orange}{0.8144} & 0.8802 & 0.9161 & 0.9374 & 0.9550 & 0.9661 & 0.9756         \\
\textbf{OC-SVM}       & {0.5532} & 0.5935 & 0.6217 & 0.6547 & 0.7019 & 0.7443 & 0.8058\\
\textbf{IF}           & {0.5475} & 0.5925 & 0.6236 & 0.6657 & 0.7188 & 0.7624 & 0.8144\\
\hline
\end{tabular}
}
\caption{Averaged detection rates for different detection algorithms under varying FDIA ratios $\%\mu$ and full measurements $\gamma = 0.00$.}
\label{table:compare_ml}
\end{table*}

The comparisons between the proposed LSTM-RDAE and the above baselines are recorded in Table \ref{table:compare_ml} assuming a full measurement condition. To have a fare comparison, we also add random input dropout mask on DAE during training. The objective of OC-SVM is to find a hyperplane that has the maximum margin separating the high dimensional data with the origin \cite{scholkopf01estimating}. The RBF kernel method is used in OC-SVM to measure the nonlinearity. Successive random hyper-separations are applied in IF \cite{ahmed19unsupervised,liu08isolation} where the abnormal data can be automatically picked out with less steps if they are \textit{rare and distinctive}. Ensemble isolation trees (\textit{iTrees}) are constructed to find the average anomaly score. The two ML models are trained solely on the normalities in the semi-supervised manner and the contamination ratios (similar to the quantile defined in AE-based methods) are tuned to have $5\%$ $FPR$ on the validation set. Principle component analysis (PCA) is applied in advance on the scaled training data to reduce the data dimension. In our example, we train the MLs under supervised awareness and it suggests that the first 26 out of 304 components can cover $99\%$ variances thus can give the best performance. In the simulation, it is assumed that the missing attributes have been imputed, i.e. $\gamma = 0.00$. The suggested hyperparameters for MLs are recorded in Table \ref{table:para_ml}.
\begin{table}[ht]
    \centering
    \begin{tabular}{lc|lc}
    \hline
        \multicolumn{2}{c|}{\textbf{OC-SVM}} & \multicolumn{2}{c}{\textbf{IF}} \\
        \hline
        Kernel method & RBF & No. of \textit{iTrees} & 200 \\
        Kernel coefficient & 0.1 & Samples per \textit{iTrees} & 256 \\
        Contamination ratio & 0.02 & Contamination ratio & 0.04 \\
        \hline
    \end{tabular}
    \caption{Hyper-parameters of the MLs}
    \label{table:para_ml}
\end{table}

As shown by Table \ref{table:compare_ml}, the performances of three deep learning methods can significantly outperform the OC-SVM and IF's. Even under intense attack strength, both the OC-SVM's and the IF's detection rates are relatively low. Firstly, as real-time load profiles are considered, different loads have different ranges and tendencies so that the obtained measurements are more complex and temporally correlated (Fig.\ref{fig:load_profile}) that may exceed the ML's competence. In the meanwhile, the first 26 components are considered through PCA which is still high for machine learning algorithms. Secondly, IF is set to be semi-supervised which leverages its separation effects on the unseen attack samples. Furthermore, the $TPR$ of the proposed method is slightly lower than LSTM-AE as the cost of considering all availability attack possibilities during training. It suggests that if there is no availability attack, LSTM-AE can be safely implemented for detection. As the FDIA becomes more intense, the assaults turn to be less stealthy to the control center and the difference between these two methods becomes negligible.

Table \ref{table:compare_dl} evaluates the $TPR$s of LSTM-AE, DAE, and the proposed LSTM-RDAE under varying FDIAs and availability attack ratios. Firstly, the detection strategy of LSTM-AE follows the method in Section \ref{sec:detection_strategy}. As it is trained without specifying missing input condition, LSTM-AE is leveraged by the large false alarm causing the lowest detection rate of the three methods. In all attack scenarios in Table \ref{table:compare_ml} and \ref{table:compare_dl}, the proposed algorithm outperforms the DAE, especially when the FDIAs are small and $\gamma$s are large (highlighted in red) which verifies that the temporal correlations extracted by the recursive LSTM cells can give more confidence on the unavailable measurements as discussed in Fig. \ref{fig:LSTM_DAE_manifold}. 

\begin{table}[ht] 
    \centering
    \begin{tabular}{c|c|c|c|c}
    \hline
        \multicolumn{5}{c}{$\%\mu = 5\%$}\\
        \hline
         Algorithms & $\gamma = 0.05$ & 0.10 & 0.15 & 0.20  \\\hline
         \textbf{LSTM-AE} & 0.8034 & 0.7592 & 0.7151 & 0.6702 \\
         \textbf{DAE} & 0.8311 & 0.8023 & 0.7746 & 0.7433\\
         \textbf{LSTM-RDAE} & 0.8565 & 0.8356 & 0.8137 & 0.7989\\
         \textbf{Improvement} & 0.0254 & 0.0333 & 0.0391 & \textcolor{red}{0.0556}\\
         \hline
         \multicolumn{5}{c}{$\%\mu = 10\%$}\\
         \hline
         \textbf{LSTM-AE} & 0.8860 & 0.8429 & 0.8107 & 0.8022 \\
         \textbf{DAE}& 0.9139 & 0.8917 & 0.8662 & 0.8423\\
         \textbf{LSTM-RDAE}& 0.9277 & 0.9168 & 0.9035 & 0.8902\\
         \textbf{Improvement} & 0.0128 & 0.0251 & 0.0373 & 0.0479\\
         \hline
         \multicolumn{5}{c}{$\%\mu = 20\%$}\\
         \hline
         \textbf{LSTM-AE} & 0.9305 & 0.9142 & 0.8979 & 0.8741 \\
         \textbf{DAE} & 0.9519 & 0.9414 & 0.9305 & 0.9187\\
         \textbf{LSTM-RDAE} & 0.9594 & 0.9536 & 0.9467 & 0.9414\\
         \textbf{Improvement} & 0.0075 & 0.0122 & 0.0162 & 0.0227\\
         \hline
    \end{tabular}
    \caption{Averaged detection rates for different detection algorithms under different FDIA and availability attack ratios. The improvement is calculated between LSTM-RDAE and DAE.}
    \label{table:compare_dl}
\end{table}


\subsubsection{Small and Successive Attacks}

\begin{figure*}[ht]
     \centering
     \begin{subfigure}[b]{0.49\textwidth}
         \centering
         \includegraphics[width=\textwidth]{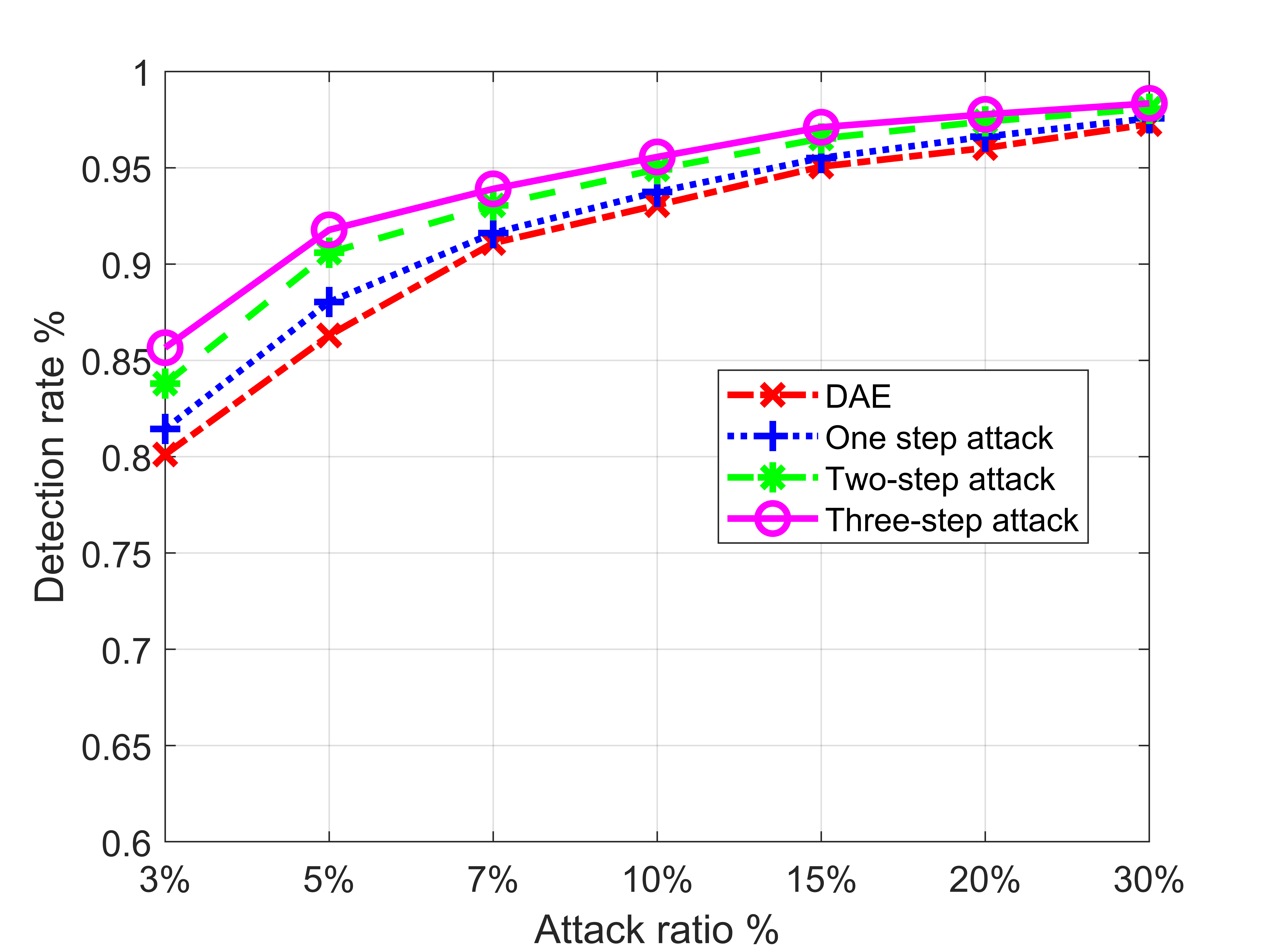}
         \caption{}
         \label{fig:y equals x}
     \end{subfigure}
     \hfill
     \begin{subfigure}[b]{0.49\textwidth}
         \centering
         \includegraphics[width=\textwidth]{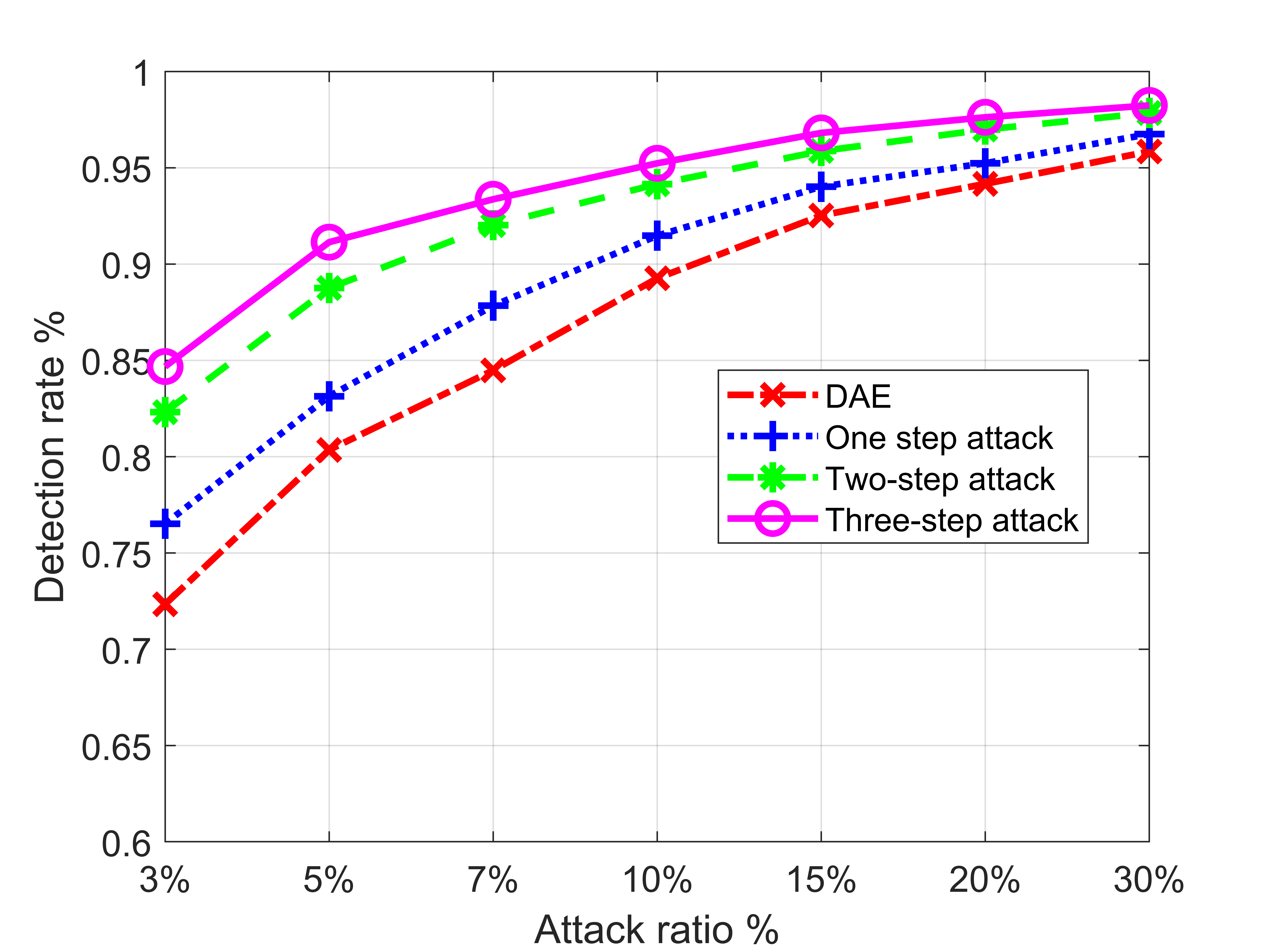}
         \caption{}
         \label{fig:three sin x}
     \end{subfigure}
     \hfill
     \begin{subfigure}[b]{0.49\textwidth}
         \centering
         \includegraphics[width=\textwidth]{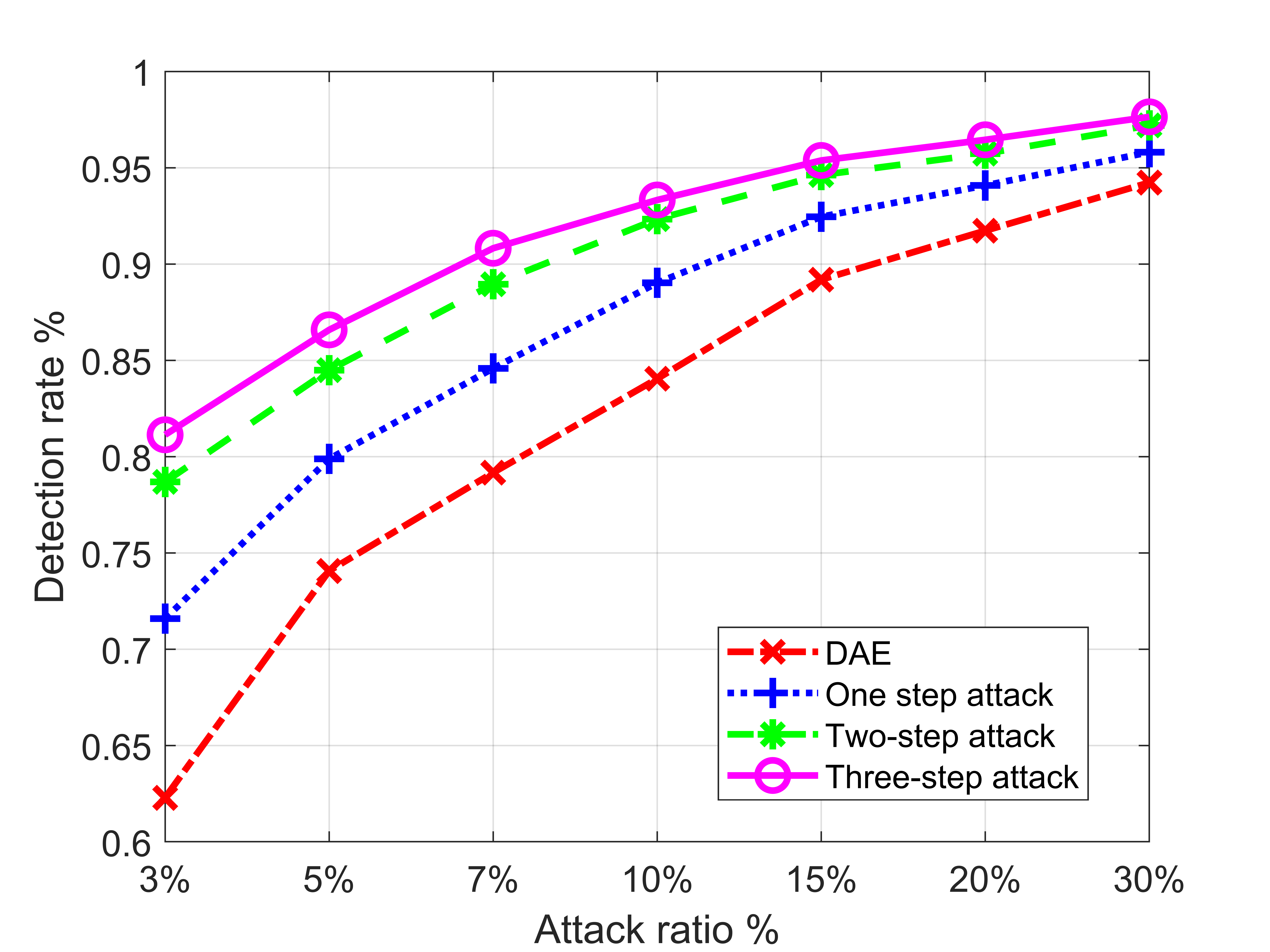}
         \caption{}
         \label{fig:five over x}
     \end{subfigure}
     \hfill
     \begin{subfigure}[b]{0.49\textwidth}
         \centering
         \includegraphics[width=\textwidth]{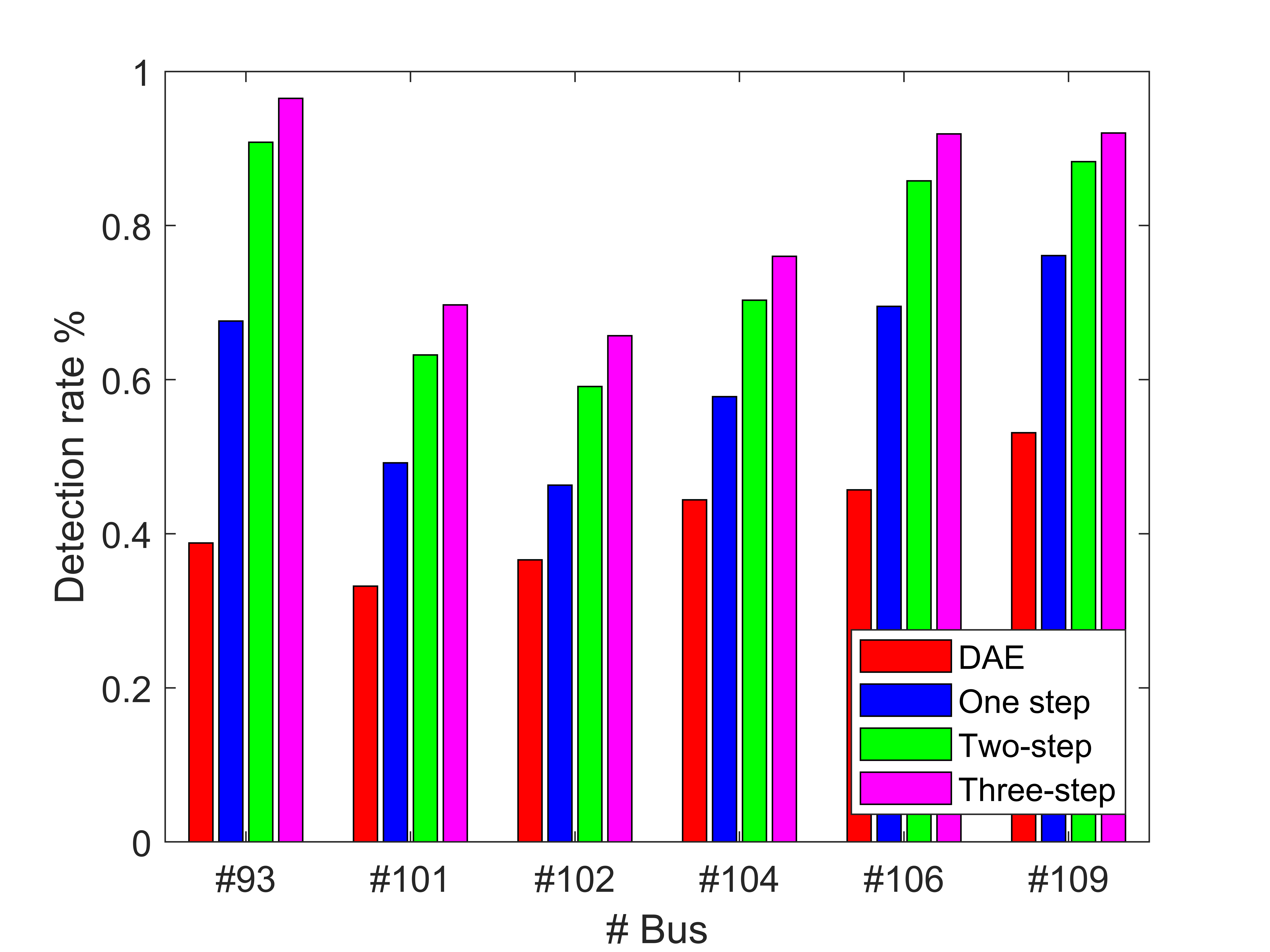}
         \caption{}
         \label{fig:five over x}
     \end{subfigure}
        \caption{Averaged detection rates under different combined FDIA-Availability attack strengths: (a) $\gamma = 0.00$; (b) $\gamma = 0.10$, (c) $\gamma = 0.20$, and (d) improvements on the selected buses with initial detection rate smaller than 80\% under 10\% FDIA attack and 20\% availability attack.}
        \label{fig:successive_attack1}
\end{figure*}

Temporal correlation is examined in this section where slight but successive attacks are considered to investigate the detection performance on the cumulative anomalies. In detail, a single attack with small attack vector may be harmless to the safety operation but their compound impacts should not be overlooked. Here, the attacks are assumed to last for at most three steps \cite{yang20bad}.

Fig.\ref{fig:successive_attack1} simulate the LSTM-RDAE under various successive combined attack strengths. As the missing ratio grows, the detection on one-shot attack becomes harder and the improvement on the detection rate introduced by successive anomalies becomes more significant, i.e. the detection rate on the three-step attacks (pink curves) are similar in all three conditions (a)-(c) no matter the missing ratio $\gamma$. Fig.\ref{fig:successive_attack1}(d) highlights the condition of selected buses with low detection rate when one-shot 10\% false data and 20\% missing ratios are injected. In majority of these buses, the detection rates when at most three steps are considered can be boosted by $10\%$ to $25\%$ compared with the one-shot attack and $20\%$ to $50\%$ compared with the DAE.    

\subsubsection{FDIAs under Targeted Availability Attacks}

In the previous sections, the measurements are lost completely at random (MCAR) while in practice, the attacker may blind a certain area of RTUs to mislead the control center. Meanwhile, it is also realistic to have missing data around a central junctional-node. To test the algorithm robustness on targeted availability attacks, the measurements are attacked according to their state degrees, thus leading to an MAR scheme. For the $i$th state, we define its \textit{attack neighborhood} $\mathcal{N}_{a}(i)$ as the set containing the measurements directly connected to the contaminated measurements:
\begin{subequations}
\begin{equation}
    \mathcal{I}_a(i) = \{j|H(j,i)\neq 0 \}
\end{equation}
\begin{equation}
    \mathcal{N}_{a}(i) = \{k|H(k,j)\neq 0,\forall j\in\mathcal{I}_a(i)   \} \setminus \mathcal{I}_a(i)
\end{equation}
\end{subequations}
where $\mathcal{I}_a(i)$ defines the contaminated measurement set due to state attack on bus $i$ and the set difference $\setminus$ ensures that the contaminated measurements in set $\mathcal{I}_a(i)$ are still available to the control center. In Table \ref{table:target_missing}, FDIAs with $\% \mu = 0.10$ on bus 93 and 94 are simulated with the following targeted availability attack settings. 
\begin{enumerate}
    \item FDIA on bus 93 with $|\mathcal{I}_a(93)|_0 = 5$ and blind on set $\mathcal{I}_d(93) = \mathcal{N}_a(93)$ which accounts for availability attack ratio $\gamma = 4.93\%$;
    \item FDIA on bus 94 with $|\mathcal{I}_a(94)|_0 = 11$ and blind on set $\mathcal{I}_d(94) =\mathcal{N}_a(94)$ which accounts for availability attack ratio $\gamma = 9.21\%$;
\end{enumerate}

\begin{figure}[ht]
  \begin{center}
  \includegraphics[width=0.5\linewidth]{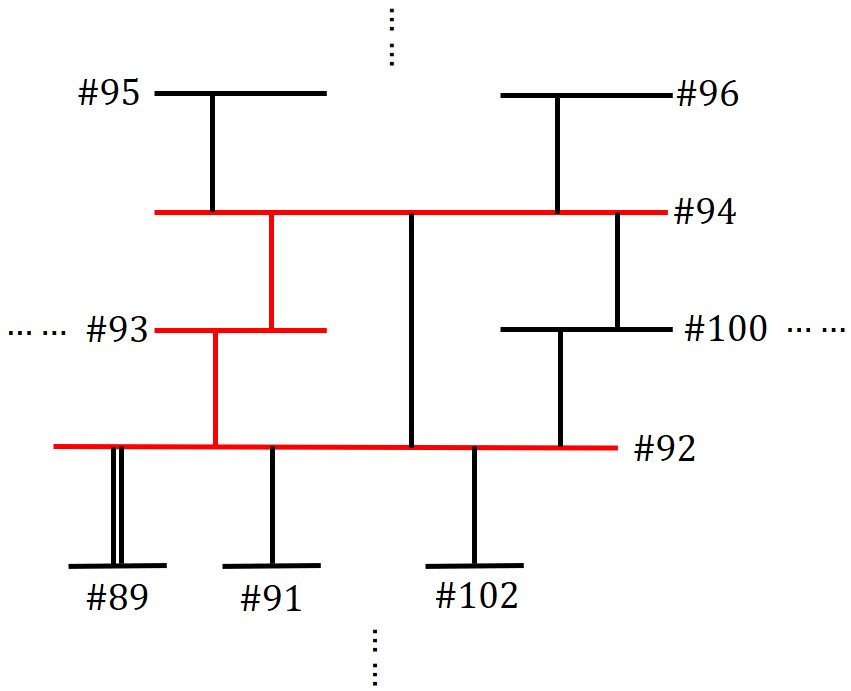}
  \caption{Illustration of proposed target availability attack on bus 93.}
  \label{fig:attack_neighbor}  
  \end{center}
\end{figure}

The mesh graph around bus 93 is illustrated by Fig.\ref{fig:attack_neighbor}. The red nodes and edges represent the contaminated measurements in set $\mathcal{I}_a(93)$ whereas the blacks represent the attack neighborhoods in set $\mathcal{N}_a(93)$ for target availability attacks. Random availability attacks are also simulated in Table \ref{table:target_missing} with the same missing ratios in the two cases. In general, masking the measurements in $\mathcal{N}_a$ will deteriorate the detection performances in both DAE and LSTM-RDAE methods as the most relevant spatial correlations are now unknown to the control center. However, the deterioration percentage on the proposed algorithm is smaller than it in DAE, which again verifies the assumption in Fig.\ref{fig:LSTM_DAE_manifold}. In the worst case, 77\% detection rate is still maintained. It is worth to note that different buses can have distinct sensitivities on the availability attack types and strengths, due to their various degrees and topological configurations.

\begin{table}[ht]
\centering
\scalebox{1}{
\begin{tabular}{c|cc|cc}
\hline
{} & \multicolumn{2}{c|}{bus 93} & \multicolumn{2}{c}{bus 94} \\
{} & target & random & target & random              \\ \hline
\textbf{DAE}            & 0.389  & 0.838  & 0.834  & 0.861\\
\textbf{LSTM-RDAE}      & 0.773  & 0.946  & 0.904  & 0.925\\
\hline
\end{tabular}
}
\caption{Detection rates under target availability attacks.}
\label{table:target_missing}
\end{table}


\subsubsection{Stealth Replay Attacks}

If the attacker is located at the control center, e.g. an internal intruder, he might have access to the previous measurements to impose the replay attack \cite{musleh19survey}. To improve the attack stealthness, we assume that the current measurement is replaced by the measurement at the same time of the previous day, e.g. ${z_a}(t)=z\left(t-t_{0}\right)$ where $t_0 = 288$. Although it has been argued that the replay attack may not be practical to the attacker in real life \cite{zhao15forecasting,zhao15short}, in general it serves as the strongest stealthy assault to bypass both BDD \eqref{eq: BDD}, \eqref{eq:imperfect_attack} and the point detection algorithm, such as DAE. Indeed, the replay attack can be seen as an FDIA with $a(t) = z_a(t)-z(t)$ and the estimated state variation can be calculated by \eqref{eq:state_estimation}. As a result, replay attack is tested as an theoretical instance of contextual anomaly. 

 The simulation results on one-shot replay attacks are illustrated in Table \ref{table:replay_attack} under different $\gamma$s. The detection rates of DAE are around 5\% which are similar to the predefined $FPR$ ($=5\%$) during the model verification stage. Thanks to the temporal exploitation property in LSTM cells, the proposed algorithm can achieve detection rate on the replay attack between 84\% and 72\% depending on the different rate of missing data.  

\begin{table}[ht]
\centering
\scalebox{1}{
\begin{tabular}{l|cccc}
\hline
\multicolumn{1}{c|}{\textbf{Strategy}} & $\gamma = 0$ & $0.05$ & $ 0.10$ & $ 0.20$               \\ \hline
\textbf{LSTM-RDAE}            & 0.8424  & 0.8106  & 0.7740  & 0.7110\\
\textbf{DAE}                 & 0.0628  & 0.0610  & 0.0567  &  0.0468          \\
\hline
\end{tabular}
}
\caption{Averaged detection rates for the previous day's replay attacks.}
\label{table:replay_attack}
\end{table}

\section{Conclusions}\label{sec:conclusion}

This paper investigates a semi-supervised detection algorithm for combined integrity-availability attack in power system where model knowledge, sample labelling, and preliminaries on the attack patterns are not required. We formulate the availability attack to further improve the stealth of FDIAs that can confuse the control option and deteriorate the false alarm ratio. The proposed LSTM-RDAE is pure data-driven and can explicitly fit on the spatiotemporal complexities in the normal measurement sequences. Moreover, a completely random dropout layer is designed after the input layer to evaluate the varying availability attack ratios. The performance of the proposed detection framework is verified under IEEE 118-bus system where real-time load profiles are employed. Sensitivity analysis is given for parameter tuning while various unseen combined attack scenarios, e.g. one-shot and successive state attack, target availability attack, and stealth replay attack are simulated during the test stage. By controlling the FPR under 5\%, the simulation results verify that the proposed LSTM-RDAE is more accurate with approximate 95\% detection rate under moderate attacks and more robust than the state-of-art deep machine learning counterparts in literature.

\printcredits

\bibliographystyle{cas-model2-names}

\bibliography{cas-refs}

\begin{thebibliography}{67}
\expandafter\ifx\csname natexlab\endcsname\relax\def\natexlab#1{#1}\fi
\providecommand{\url}[1]{\texttt{#1}}
\providecommand{\href}[2]{#2}
\providecommand{\path}[1]{#1}
\providecommand{\DOIprefix}{doi:}
\providecommand{\ArXivprefix}{arXiv:}
\providecommand{\URLprefix}{URL: }
\providecommand{\Pubmedprefix}{pmid:}
\providecommand{\doi}[1]{\href{http://dx.doi.org/#1}{\path{#1}}}
\providecommand{\Pubmed}[1]{\href{pmid:#1}{\path{#1}}}
\providecommand{\bibinfo}[2]{#2}
\ifx\xfnm\relax \def\xfnm[#1]{\unskip,\space#1}\fi
\bibitem[{Abur and Exposito(2004)}]{abur04power}
\bibinfo{author}{Abur, A.}, \bibinfo{author}{Exposito, A.G.},
  \bibinfo{year}{2004}.
\newblock \bibinfo{title}{Power system state estimation: theory and
  implementation}.
\newblock \bibinfo{publisher}{CRC press}.
\bibitem[{{Ahmed} et~al.(2019){Ahmed}, {Lee}, {Hyun} and
  {Koo}}]{ahmed19unsupervised}
\bibinfo{author}{{Ahmed}, S.}, \bibinfo{author}{{Lee}, Y.},
  \bibinfo{author}{{Hyun}, S.}, \bibinfo{author}{{Koo}, I.},
  \bibinfo{year}{2019}.
\newblock \bibinfo{title}{Unsupervised machine learning-based detection of
  covert data integrity assault in smart grid networks utilizing isolation
  forest}.
\newblock \bibinfo{journal}{IEEE Transactions on Information Forensics and
  Security} \bibinfo{volume}{14}, \bibinfo{pages}{2765--2777}.
\bibitem[{Aslani et~al.(2022)Aslani, Faraji, Hashemi-Dezaki and
  Ketabi}]{aslani2022novel}
\bibinfo{author}{Aslani, M.}, \bibinfo{author}{Faraji, J.},
  \bibinfo{author}{Hashemi-Dezaki, H.}, \bibinfo{author}{Ketabi, A.},
  \bibinfo{year}{2022}.
\newblock \bibinfo{title}{A novel clustering-based method for reliability
  assessment of cyber-physical microgrids considering cyber interdependencies
  and information transmission errors}.
\newblock \bibinfo{journal}{Applied Energy} \bibinfo{volume}{315},
  \bibinfo{pages}{119032}.
\bibitem[{Bi et~al.(2022)Bi, Luo, He, Liang, Meng and Sun}]{bi2022false}
\bibinfo{author}{Bi, J.}, \bibinfo{author}{Luo, F.}, \bibinfo{author}{He, S.},
  \bibinfo{author}{Liang, G.}, \bibinfo{author}{Meng, W.},
  \bibinfo{author}{Sun, M.}, \bibinfo{year}{2022}.
\newblock \bibinfo{title}{False data injection-and propagation-aware game
  theoretical approach for microgrids}.
\newblock \bibinfo{journal}{IEEE Transactions on Smart Grid} .
\bibitem[{Chalapathy and Chawla(2019)}]{chalapathy19deep}
\bibinfo{author}{Chalapathy, R.}, \bibinfo{author}{Chawla, S.},
  \bibinfo{year}{2019}.
\newblock \bibinfo{title}{Deep learning for anomaly detection: A survey}.
\newblock \bibinfo{journal}{arXiv preprint arXiv:1901.03407} .
\bibitem[{Chandola et~al.(2009)Chandola, Banerjee and
  Kumar}]{chandola09anomaly}
\bibinfo{author}{Chandola, V.}, \bibinfo{author}{Banerjee, A.},
  \bibinfo{author}{Kumar, V.}, \bibinfo{year}{2009}.
\newblock \bibinfo{title}{Anomaly detection: A survey}.
\newblock \bibinfo{journal}{ACM computing surveys (CSUR)} \bibinfo{volume}{41},
  \bibinfo{pages}{1--58}.
\bibitem[{Chen et~al.(2020)Chen, Cui, Fang, Ren and Chen}]{chen2020load}
\bibinfo{author}{Chen, C.}, \bibinfo{author}{Cui, M.}, \bibinfo{author}{Fang,
  X.}, \bibinfo{author}{Ren, B.}, \bibinfo{author}{Chen, Y.},
  \bibinfo{year}{2020}.
\newblock \bibinfo{title}{Load altering attack-tolerant defense strategy for
  load frequency control system}.
\newblock \bibinfo{journal}{Applied Energy} \bibinfo{volume}{280},
  \bibinfo{pages}{116015}.
\bibitem[{Chen et~al.(2022)Chen, Wang, Cui, Zhao, Bi, Chen and
  Zhang}]{chen2022data}
\bibinfo{author}{Chen, C.}, \bibinfo{author}{Wang, Y.}, \bibinfo{author}{Cui,
  M.}, \bibinfo{author}{Zhao, J.}, \bibinfo{author}{Bi, W.},
  \bibinfo{author}{Chen, Y.}, \bibinfo{author}{Zhang, X.},
  \bibinfo{year}{2022}.
\newblock \bibinfo{title}{Data-driven detection of stealthy false data
  injection attack against power system state estimation}.
\newblock \bibinfo{journal}{IEEE Transactions on Industrial Informatics} .
\bibitem[{Chen et~al.(2017)Chen, Sathe, Aggarwal and Turaga}]{chen17outlier}
\bibinfo{author}{Chen, J.}, \bibinfo{author}{Sathe, S.},
  \bibinfo{author}{Aggarwal, C.}, \bibinfo{author}{Turaga, D.},
  \bibinfo{year}{2017}.
\newblock \bibinfo{title}{Outlier detection with autoencoder ensembles}, in:
  \bibinfo{booktitle}{Proceedings of the 2017 SIAM international conference on
  data mining}, \bibinfo{organization}{SIAM}. pp. \bibinfo{pages}{90--98}.
\bibitem[{Choeum and Choi(2021)}]{choeum2021trilevel}
\bibinfo{author}{Choeum, D.}, \bibinfo{author}{Choi, D.H.},
  \bibinfo{year}{2021}.
\newblock \bibinfo{title}{Trilevel smart meter hardening strategy for
  mitigating cyber attacks against volt/var optimization in smart power
  distribution systems}.
\newblock \bibinfo{journal}{Applied Energy} \bibinfo{volume}{304},
  \bibinfo{pages}{117710}.
\bibitem[{Chu et~al.(2020)Chu, Kosut and Sankar}]{chu20detecting}
\bibinfo{author}{Chu, Z.}, \bibinfo{author}{Kosut, O.},
  \bibinfo{author}{Sankar, L.}, \bibinfo{year}{2020}.
\newblock \bibinfo{title}{Detecting load redistribution attacks via support
  vector models}.
\newblock \bibinfo{journal}{arXiv preprint arXiv:2003.06543} .
\bibitem[{{Cui} et~al.(2020){Cui}, {Wang} and {Chen}}]{cui20flexible}
\bibinfo{author}{{Cui}, M.}, \bibinfo{author}{{Wang}, J.},
  \bibinfo{author}{{Chen}, B.}, \bibinfo{year}{2020}.
\newblock \bibinfo{title}{Flexible machine learning-based cyberattack detection
  using spatiotemporal patterns for distribution systems}.
\newblock \bibinfo{journal}{IEEE Transactions on Smart Grid}
  \bibinfo{volume}{11}, \bibinfo{pages}{1805--1808}.
\bibitem[{{Foroutan} and {Salmasi}(2017)}]{foroutan17detection}
\bibinfo{author}{{Foroutan}, S.A.}, \bibinfo{author}{{Salmasi}, F.R.},
  \bibinfo{year}{2017}.
\newblock \bibinfo{title}{Detection of false data injection attacks against
  state estimation in smart grids based on a mixture gaussian distribution
  learning method}.
\newblock \bibinfo{journal}{IET Cyber-Physical Systems: Theory Applications}
  \bibinfo{volume}{2}, \bibinfo{pages}{161--171}.
\bibitem[{Gal and Ghahramani(2016)}]{gal16theoretically}
\bibinfo{author}{Gal, Y.}, \bibinfo{author}{Ghahramani, Z.},
  \bibinfo{year}{2016}.
\newblock \bibinfo{title}{A theoretically grounded application of dropout in
  recurrent neural networks}, in: \bibinfo{booktitle}{Advances in neural
  information processing systems}, pp. \bibinfo{pages}{1019--1027}.
\bibitem[{Gao et~al.(2022)Gao, Lei, Wei, Liu and Wang}]{gao2022novel}
\bibinfo{author}{Gao, S.}, \bibinfo{author}{Lei, J.}, \bibinfo{author}{Wei,
  X.}, \bibinfo{author}{Liu, Y.}, \bibinfo{author}{Wang, T.},
  \bibinfo{year}{2022}.
\newblock \bibinfo{title}{A novel bilevel false data injection attack model
  based on pre-and post-dispatch}.
\newblock \bibinfo{journal}{IEEE Transactions on Smart Grid}
  \bibinfo{volume}{13}, \bibinfo{pages}{2487--2490}.
\bibitem[{Goodfellow et~al.(2016)Goodfellow, Bengio and
  Courville}]{goodfellow2016deep}
\bibinfo{author}{Goodfellow, I.}, \bibinfo{author}{Bengio, Y.},
  \bibinfo{author}{Courville, A.}, \bibinfo{year}{2016}.
\newblock \bibinfo{title}{Deep learning}.
\newblock \bibinfo{publisher}{MIT press}.
\bibitem[{{He} et~al.(2017){He}, {Mendis} and {Wei}}]{he17real}
\bibinfo{author}{{He}, Y.}, \bibinfo{author}{{Mendis}, G.J.},
  \bibinfo{author}{{Wei}, J.}, \bibinfo{year}{2017}.
\newblock \bibinfo{title}{Real-time detection of false data injection attacks
  in smart grid: A deep learning-based intelligent mechanism}.
\newblock \bibinfo{journal}{IEEE Transactions on Smart Grid}
  \bibinfo{volume}{8}, \bibinfo{pages}{2505--2516}.
\bibitem[{Hinton and Salakhutdinov(2006)}]{hinton06reducing}
\bibinfo{author}{Hinton, G.E.}, \bibinfo{author}{Salakhutdinov, R.R.},
  \bibinfo{year}{2006}.
\newblock \bibinfo{title}{Reducing the dimensionality of data with neural
  networks}.
\newblock \bibinfo{journal}{science} \bibinfo{volume}{313},
  \bibinfo{pages}{504--507}.
\bibitem[{Hirth et~al.(2019)Hirth, Muhlenohordt, Schlecht and
  Weibezhhn}]{hirth19time}
\bibinfo{author}{Hirth, L.}, \bibinfo{author}{Muhlenohordt, J.},
  \bibinfo{author}{Schlecht, I.}, \bibinfo{author}{Weibezhhn, J.},
  \bibinfo{year}{2019}.
\newblock \bibinfo{title}{Time series data}.
\newblock \URLprefix
  \url{https://data.open-power-system-data.org/time_series/2019-06-05}.
\bibitem[{Hochreiter and Schmidhuber(1997)}]{hochreiter97long}
\bibinfo{author}{Hochreiter, S.}, \bibinfo{author}{Schmidhuber, J.},
  \bibinfo{year}{1997}.
\newblock \bibinfo{title}{Long short-term memory}.
\newblock \bibinfo{journal}{Neural computation} \bibinfo{volume}{9},
  \bibinfo{pages}{1735--1780}.
\bibitem[{Hug and Giampapa(2012)}]{hug12vulnerability}
\bibinfo{author}{Hug, G.}, \bibinfo{author}{Giampapa, J.A.},
  \bibinfo{year}{2012}.
\newblock \bibinfo{title}{Vulnerability assessment of ac state estimation with
  respect to false data injection cyber-attacks}.
\newblock \bibinfo{journal}{IEEE Transactions on smart grid}
  \bibinfo{volume}{3}, \bibinfo{pages}{1362--1370}.
\bibitem[{Ibrahim et~al.(2020)Ibrahim, Dong and Yang}]{ibrahim2020machine}
\bibinfo{author}{Ibrahim, M.S.}, \bibinfo{author}{Dong, W.},
  \bibinfo{author}{Yang, Q.}, \bibinfo{year}{2020}.
\newblock \bibinfo{title}{Machine learning driven smart electric power systems:
  Current trends and new perspectives}.
\newblock \bibinfo{journal}{Applied Energy} \bibinfo{volume}{272},
  \bibinfo{pages}{115237}.
\bibitem[{{Jindal} et~al.(2016){Jindal}, {Dua}, {Kaur}, {Singh}, {Kumar} and
  {Mishra}}]{jindal16decision}
\bibinfo{author}{{Jindal}, A.}, \bibinfo{author}{{Dua}, A.},
  \bibinfo{author}{{Kaur}, K.}, \bibinfo{author}{{Singh}, M.},
  \bibinfo{author}{{Kumar}, N.}, \bibinfo{author}{{Mishra}, S.},
  \bibinfo{year}{2016}.
\newblock \bibinfo{title}{Decision tree and svm-based data analytics for theft
  detection in smart grid}.
\newblock \bibinfo{journal}{IEEE Transactions on Industrial Informatics}
  \bibinfo{volume}{12}, \bibinfo{pages}{1005--1016}.
\bibitem[{Kieu et~al.(2019)Kieu, Yang, Guo and Jensen}]{kieu19outlier}
\bibinfo{author}{Kieu, T.}, \bibinfo{author}{Yang, B.}, \bibinfo{author}{Guo,
  C.}, \bibinfo{author}{Jensen, C.S.}, \bibinfo{year}{2019}.
\newblock \bibinfo{title}{Outlier detection for time series with recurrent
  autoencoder ensembles.}, in: \bibinfo{booktitle}{IJCAI}, pp.
  \bibinfo{pages}{2725--2732}.
\bibitem[{{Kosut} et~al.(2011){Kosut}, {Jia}, {Thomas} and
  {Tong}}]{kosut11malicious}
\bibinfo{author}{{Kosut}, O.}, \bibinfo{author}{{Jia}, L.},
  \bibinfo{author}{{Thomas}, R.J.}, \bibinfo{author}{{Tong}, L.},
  \bibinfo{year}{2011}.
\newblock \bibinfo{title}{Malicious data attacks on the smart grid}.
\newblock \bibinfo{journal}{IEEE Transactions on Smart Grid}
  \bibinfo{volume}{2}, \bibinfo{pages}{645--658}.
\bibitem[{Lai et~al.(2019)Lai, Illindala and Subramaniam}]{lai2019tri}
\bibinfo{author}{Lai, K.}, \bibinfo{author}{Illindala, M.},
  \bibinfo{author}{Subramaniam, K.}, \bibinfo{year}{2019}.
\newblock \bibinfo{title}{A tri-level optimization model to mitigate
  coordinated attacks on electric power systems in a cyber-physical
  environment}.
\newblock \bibinfo{journal}{Applied energy} \bibinfo{volume}{235},
  \bibinfo{pages}{204--218}.
\bibitem[{Lee et~al.(2016)Lee, Assante and Conway}]{lee16analysis}
\bibinfo{author}{Lee, R.M.}, \bibinfo{author}{Assante, M.J.},
  \bibinfo{author}{Conway, T.}, \bibinfo{year}{2016}.
\newblock \bibinfo{title}{Analysis of the cyber attack on the ukrainian power
  grid}.
\newblock \bibinfo{journal}{Electricity Information Sharing and Analysis Center
  (E-ISAC)} \bibinfo{volume}{388}.
\bibitem[{{Lin} and {Wang}(2020)}]{lin20probabilistic}
\bibinfo{author}{{Lin}, Y.}, \bibinfo{author}{{Wang}, J.},
  \bibinfo{year}{2020}.
\newblock \bibinfo{title}{Probabilistic deep autoencoder for power system
  measurement outlier detection and reconstruction}.
\newblock \bibinfo{journal}{IEEE Transactions on Smart Grid}
  \bibinfo{volume}{11}, \bibinfo{pages}{1796--1798}.
\bibitem[{{Liu} et~al.(2008){Liu}, {Ting} and {Zhou}}]{liu08isolation}
\bibinfo{author}{{Liu}, F.T.}, \bibinfo{author}{{Ting}, K.M.},
  \bibinfo{author}{{Zhou}, Z.}, \bibinfo{year}{2008}.
\newblock \bibinfo{title}{Isolation forest}, in: \bibinfo{booktitle}{2008
  Eighth IEEE International Conference on Data Mining}, pp.
  \bibinfo{pages}{413--422}.
\bibitem[{{Liu} et~al.(2016){Liu}, {Li}, {Liu} and {Li}}]{liu16masking}
\bibinfo{author}{{Liu}, X.}, \bibinfo{author}{{Li}, Z.},
  \bibinfo{author}{{Liu}, X.}, \bibinfo{author}{{Li}, Z.},
  \bibinfo{year}{2016}.
\newblock \bibinfo{title}{Masking transmission line outages via false data
  injection attacks}.
\newblock \bibinfo{journal}{IEEE Transactions on Information Forensics and
  Security} \bibinfo{volume}{11}, \bibinfo{pages}{1592--1602}.
\bibitem[{Liu et~al.(2011)Liu, Ning and Reiter}]{liu11false}
\bibinfo{author}{Liu, Y.}, \bibinfo{author}{Ning, P.}, \bibinfo{author}{Reiter,
  M.K.}, \bibinfo{year}{2011}.
\newblock \bibinfo{title}{False data injection attacks against state estimation
  in electric power grids}.
\newblock \bibinfo{journal}{ACM Transactions on Information and System Security
  (TISSEC)} \bibinfo{volume}{14}, \bibinfo{pages}{1--33}.
\bibitem[{{Mo} and {Sinopoli}(2009)}]{mo09secure}
\bibinfo{author}{{Mo}, Y.}, \bibinfo{author}{{Sinopoli}, B.},
  \bibinfo{year}{2009}.
\newblock \bibinfo{title}{Secure control against replay attacks}, in:
  \bibinfo{booktitle}{2009 47th Annual Allerton Conference on Communication,
  Control, and Computing (Allerton)}, pp. \bibinfo{pages}{911--918}.
\bibitem[{Moayyed et~al.(2021)Moayyed, Mohammadpourfard, Konstantinou,
  Moradzadeh, Mohammadi-Ivatloo and Aguiar}]{moayyed2021image}
\bibinfo{author}{Moayyed, H.}, \bibinfo{author}{Mohammadpourfard, M.},
  \bibinfo{author}{Konstantinou, C.}, \bibinfo{author}{Moradzadeh, A.},
  \bibinfo{author}{Mohammadi-Ivatloo, B.}, \bibinfo{author}{Aguiar, A.P.},
  \bibinfo{year}{2021}.
\newblock \bibinfo{title}{Image processing based approach for false data
  injection attacks detection in power systems}.
\newblock \bibinfo{journal}{IEEE Access} \bibinfo{volume}{10},
  \bibinfo{pages}{12412--12420}.
\bibitem[{Musleh et~al.(2020)Musleh, Chen and Dong}]{musleh19survey}
\bibinfo{author}{Musleh, A.S.}, \bibinfo{author}{Chen, G.},
  \bibinfo{author}{Dong, Z.Y.}, \bibinfo{year}{2020}.
\newblock \bibinfo{title}{A survey on the detection algorithms for false data
  injection attacks in smart grids}.
\newblock \bibinfo{journal}{IEEE Transactions on Smart Grid}
  \bibinfo{volume}{11}, \bibinfo{pages}{2218--2234}.
\bibitem[{{Ozay} et~al.(2016){Ozay}, {Esnaola}, {Yarman Vural}, {Kulkarni} and
  {Poor}}]{ozay16machine}
\bibinfo{author}{{Ozay}, M.}, \bibinfo{author}{{Esnaola}, I.},
  \bibinfo{author}{{Yarman Vural}, F.T.}, \bibinfo{author}{{Kulkarni}, S.R.},
  \bibinfo{author}{{Poor}, H.V.}, \bibinfo{year}{2016}.
\newblock \bibinfo{title}{Machine learning methods for attack detection in the
  smart grid}.
\newblock \bibinfo{journal}{IEEE Transactions on Neural Networks and Learning
  Systems} \bibinfo{volume}{27}, \bibinfo{pages}{1773--1786}.
\bibitem[{{Pan} et~al.(2019){Pan}, {Teixeira}, {Cvetkovic} and
  {Palensky}}]{pan19cyber}
\bibinfo{author}{{Pan}, K.}, \bibinfo{author}{{Teixeira}, A.},
  \bibinfo{author}{{Cvetkovic}, M.}, \bibinfo{author}{{Palensky}, P.},
  \bibinfo{year}{2019}.
\newblock \bibinfo{title}{Cyber risk analysis of combined data attacks against
  power system state estimation}.
\newblock \bibinfo{journal}{IEEE Transactions on Smart Grid}
  \bibinfo{volume}{10}, \bibinfo{pages}{3044--3056}.
\bibitem[{{Pan} et~al.(2016){Pan}, {Teixeira}, {Cvetkovic} and
  {Palensky}}]{pan16combined}
\bibinfo{author}{{Pan}, K.}, \bibinfo{author}{{Teixeira}, A.M.H.},
  \bibinfo{author}{{Cvetkovic}, M.}, \bibinfo{author}{{Palensky}, P.},
  \bibinfo{year}{2016}.
\newblock \bibinfo{title}{Combined data integrity and availability attacks on
  state estimation in cyber-physical power grids}, in: \bibinfo{booktitle}{2016
  IEEE International Conference on Smart Grid Communications (SmartGridComm)},
  pp. \bibinfo{pages}{271--277}.
\bibitem[{Pang et~al.(2020)Pang, Shen, Cao and Hengel}]{pang20deep}
\bibinfo{author}{Pang, G.}, \bibinfo{author}{Shen, C.}, \bibinfo{author}{Cao,
  L.}, \bibinfo{author}{Hengel, A.v.d.}, \bibinfo{year}{2020}.
\newblock \bibinfo{title}{Deep learning for anomaly detection: A review}.
\newblock \bibinfo{journal}{arXiv preprint arXiv:2007.02500} .
\bibitem[{{Qiu} et~al.(2020){Qiu}, {Tang}, {Zhu}, {Wang}, {Liu} and
  {Yao}}]{qiu20detection}
\bibinfo{author}{{Qiu}, W.}, \bibinfo{author}{{Tang}, Q.},
  \bibinfo{author}{{Zhu}, K.}, \bibinfo{author}{{Wang}, W.},
  \bibinfo{author}{{Liu}, Y.}, \bibinfo{author}{{Yao}, W.},
  \bibinfo{year}{2020}.
\newblock \bibinfo{title}{Detection of synchrophasor false data injection
  attack using feature interactive network}.
\newblock \bibinfo{journal}{IEEE Transactions on Smart Grid} ,
  \bibinfo{pages}{1--1}.
\bibitem[{Rahman and Mohsenian-Rad(2013)}]{rahman13false}
\bibinfo{author}{Rahman, M.A.}, \bibinfo{author}{Mohsenian-Rad, H.},
  \bibinfo{year}{2013}.
\newblock \bibinfo{title}{False data injection attacks against nonlinear state
  estimation in smart power grids}, in: \bibinfo{booktitle}{2013 IEEE Power \&
  Energy Society General Meeting}, \bibinfo{organization}{IEEE}. pp.
  \bibinfo{pages}{1--5}.
\bibitem[{Rubin(1976)}]{rubin76inference}
\bibinfo{author}{Rubin, D.B.}, \bibinfo{year}{1976}.
\newblock \bibinfo{title}{Inference and missing data}.
\newblock \bibinfo{journal}{Biometrika} \bibinfo{volume}{63},
  \bibinfo{pages}{581--592}.
\bibitem[{Ruff et~al.(2018)Ruff, Vandermeulen, Goernitz, Deecke, Siddiqui,
  Binder, M{\"u}ller and Kloft}]{ruff18deep}
\bibinfo{author}{Ruff, L.}, \bibinfo{author}{Vandermeulen, R.},
  \bibinfo{author}{Goernitz, N.}, \bibinfo{author}{Deecke, L.},
  \bibinfo{author}{Siddiqui, S.A.}, \bibinfo{author}{Binder, A.},
  \bibinfo{author}{M{\"u}ller, E.}, \bibinfo{author}{Kloft, M.},
  \bibinfo{year}{2018}.
\newblock \bibinfo{title}{Deep one-class classification}, in:
  \bibinfo{booktitle}{International conference on machine learning}, pp.
  \bibinfo{pages}{4393--4402}.
\bibitem[{Saadat et~al.(1999)}]{saadat99power}
\bibinfo{author}{Saadat, H.}, et~al., \bibinfo{year}{1999}.
\newblock \bibinfo{title}{Power system analysis}. volume~\bibinfo{volume}{2}.
\newblock \bibinfo{publisher}{McGraw-Hill}.
\bibitem[{Sayghe et~al.(2020)Sayghe, Hu, Zografopoulos, Liu, Dutta, Jin and
  Konstantinou}]{sayghe20survey}
\bibinfo{author}{Sayghe, A.}, \bibinfo{author}{Hu, Y.},
  \bibinfo{author}{Zografopoulos, I.}, \bibinfo{author}{Liu, X.},
  \bibinfo{author}{Dutta, R.G.}, \bibinfo{author}{Jin, Y.},
  \bibinfo{author}{Konstantinou, C.}, \bibinfo{year}{2020}.
\newblock \bibinfo{title}{A survey of machine learning methods for detecting
  false data injection attacks in power systems}.
\newblock \bibinfo{journal}{arXiv preprint arXiv:2008.06926} .
\bibitem[{Sch{\"o}lkopf et~al.(2001)Sch{\"o}lkopf, Platt, Shawe-Taylor, Smola
  and Williamson}]{scholkopf01estimating}
\bibinfo{author}{Sch{\"o}lkopf, B.}, \bibinfo{author}{Platt, J.C.},
  \bibinfo{author}{Shawe-Taylor, J.}, \bibinfo{author}{Smola, A.J.},
  \bibinfo{author}{Williamson, R.C.}, \bibinfo{year}{2001}.
\newblock \bibinfo{title}{Estimating the support of a high-dimensional
  distribution}.
\newblock \bibinfo{journal}{Neural computation} \bibinfo{volume}{13},
  \bibinfo{pages}{1443--1471}.
\bibitem[{{Singh} et~al.(2018){Singh}, {Khanna}, {Bose}, {Panigrahi} and
  {Joshi}}]{singh18joint}
\bibinfo{author}{{Singh}, S.K.}, \bibinfo{author}{{Khanna}, K.},
  \bibinfo{author}{{Bose}, R.}, \bibinfo{author}{{Panigrahi}, B.K.},
  \bibinfo{author}{{Joshi}, A.}, \bibinfo{year}{2018}.
\newblock \bibinfo{title}{Joint-transformation-based detection of false data
  injection attacks in smart grid}.
\newblock \bibinfo{journal}{IEEE Transactions on Industrial Informatics}
  \bibinfo{volume}{14}, \bibinfo{pages}{89--97}.
\bibitem[{Tan et~al.(2022)Tan, Xie, Guerrero and Vasquez}]{tan2022false}
\bibinfo{author}{Tan, S.}, \bibinfo{author}{Xie, P.},
  \bibinfo{author}{Guerrero, J.M.}, \bibinfo{author}{Vasquez, J.C.},
  \bibinfo{year}{2022}.
\newblock \bibinfo{title}{False data injection cyber-attacks detection for
  multiple dc microgrid clusters}.
\newblock \bibinfo{journal}{Applied Energy} \bibinfo{volume}{310},
  \bibinfo{pages}{118425}.
\bibitem[{Teng et~al.(2016)Teng, Aunedi and Strbac}]{teng2016benefits}
\bibinfo{author}{Teng, F.}, \bibinfo{author}{Aunedi, M.},
  \bibinfo{author}{Strbac, G.}, \bibinfo{year}{2016}.
\newblock \bibinfo{title}{Benefits of flexibility from smart electrified
  transportation and heating in the future uk electricity system}.
\newblock \bibinfo{journal}{Applied energy} \bibinfo{volume}{167},
  \bibinfo{pages}{420--431}.
\bibitem[{Tian et~al.(2020)Tian, Wang, Li, Shang and Cao}]{tian20coordinated}
\bibinfo{author}{Tian, J.}, \bibinfo{author}{Wang, B.}, \bibinfo{author}{Li,
  T.}, \bibinfo{author}{Shang, F.}, \bibinfo{author}{Cao, K.},
  \bibinfo{year}{2020}.
\newblock \bibinfo{title}{Coordinated cyber-physical attacks considering dos
  attacks in power systems}.
\newblock \bibinfo{journal}{International Journal of Robust and Nonlinear
  Control} \bibinfo{volume}{30}, \bibinfo{pages}{4345--4358}.
\bibitem[{Vincent et~al.(2008)Vincent, Larochelle, Bengio and
  Manzagol}]{vincent08extracting}
\bibinfo{author}{Vincent, P.}, \bibinfo{author}{Larochelle, H.},
  \bibinfo{author}{Bengio, Y.}, \bibinfo{author}{Manzagol, P.A.},
  \bibinfo{year}{2008}.
\newblock \bibinfo{title}{Extracting and composing robust features with
  denoising autoencoders}, in: \bibinfo{booktitle}{Proceedings of the 25th
  international conference on Machine learning}, pp.
  \bibinfo{pages}{1096--1103}.
\bibitem[{Vincent et~al.(2010)Vincent, Larochelle, Lajoie, Bengio, Manzagol and
  Bottou}]{vincent10stacked}
\bibinfo{author}{Vincent, P.}, \bibinfo{author}{Larochelle, H.},
  \bibinfo{author}{Lajoie, I.}, \bibinfo{author}{Bengio, Y.},
  \bibinfo{author}{Manzagol, P.A.}, \bibinfo{author}{Bottou, L.},
  \bibinfo{year}{2010}.
\newblock \bibinfo{title}{Stacked denoising autoencoders: Learning useful
  representations in a deep network with a local denoising criterion.}
\newblock \bibinfo{journal}{Journal of machine learning research}
  \bibinfo{volume}{11}.
\bibitem[{Wang et~al.(2020a)Wang, Tindemans, Pan and
  Palensky}]{wang20detection}
\bibinfo{author}{Wang, C.}, \bibinfo{author}{Tindemans, S.},
  \bibinfo{author}{Pan, K.}, \bibinfo{author}{Palensky, P.},
  \bibinfo{year}{2020}a.
\newblock \bibinfo{title}{Detection of false data injection attacks using the
  autoencoder approach}.
\newblock \bibinfo{journal}{arXiv preprint arXiv:2003.02229} .
\bibitem[{Wang et~al.(2020b)Wang, Wen, Xu, Zhou, Peng and
  Liu}]{wang20operating}
\bibinfo{author}{Wang, H.}, \bibinfo{author}{Wen, X.}, \bibinfo{author}{Xu,
  Y.}, \bibinfo{author}{Zhou, B.}, \bibinfo{author}{Peng, J.C.},
  \bibinfo{author}{Liu, W.}, \bibinfo{year}{2020}b.
\newblock \bibinfo{title}{Operating state reconstruction in cyber physical
  smart grid for automatic attack filtering}.
\newblock \bibinfo{journal}{IEEE Transactions on Industrial Informatics} .
\bibitem[{{Wang} et~al.(2020){Wang}, {Zhou} and {Jin}}]{wang20physics}
\bibinfo{author}{{Wang}, L.}, \bibinfo{author}{{Zhou}, Q.},
  \bibinfo{author}{{Jin}, S.}, \bibinfo{year}{2020}.
\newblock \bibinfo{title}{Physics-guided deep learning for power system state
  estimation}.
\newblock \bibinfo{journal}{Journal of Modern Power Systems and Clean Energy}
  \bibinfo{volume}{8}, \bibinfo{pages}{607--615}.
\bibitem[{Wu et~al.(2020)Wu, Xue, Wang, Chung, Wang, Peng and
  Yang}]{wu20extreme}
\bibinfo{author}{Wu, T.}, \bibinfo{author}{Xue, W.}, \bibinfo{author}{Wang,
  H.}, \bibinfo{author}{Chung, C.}, \bibinfo{author}{Wang, G.},
  \bibinfo{author}{Peng, J.}, \bibinfo{author}{Yang, Q.}, \bibinfo{year}{2020}.
\newblock \bibinfo{title}{Extreme learning machine-based state reconstruction
  for automatic attack filtering in cyber physical power system}.
\newblock \bibinfo{journal}{IEEE Transactions on Industrial Informatics} .
\bibitem[{Xiong et~al.(2022)Xiong, Hu, Sun, Hao, Li and
  Lin}]{xiong2022detection}
\bibinfo{author}{Xiong, X.}, \bibinfo{author}{Hu, S.}, \bibinfo{author}{Sun,
  D.}, \bibinfo{author}{Hao, S.}, \bibinfo{author}{Li, H.},
  \bibinfo{author}{Lin, G.}, \bibinfo{year}{2022}.
\newblock \bibinfo{title}{Detection of false data injection attack in power
  information physical system based on svm--gab algorithm}.
\newblock \bibinfo{journal}{Energy Reports} \bibinfo{volume}{8},
  \bibinfo{pages}{1156--1164}.
\bibitem[{Yang et~al.(2020a)Yang, Zhang, Xiang, Liu, Liu, Han and
  Teng}]{yang20lstm}
\bibinfo{author}{Yang, J.}, \bibinfo{author}{Zhang, S.},
  \bibinfo{author}{Xiang, Y.}, \bibinfo{author}{Liu, J.}, \bibinfo{author}{Liu,
  J.}, \bibinfo{author}{Han, X.}, \bibinfo{author}{Teng, F.},
  \bibinfo{year}{2020}a.
\newblock \bibinfo{title}{Lstm auto-encoder based representative scenario
  generation method for hybrid hydro-pv power system}.
\newblock \bibinfo{journal}{IET Generation, Transmission \& Distribution} .
\bibitem[{{Yang} et~al.(2014){Yang}, {Yang}, {Yu}, {An}, {Zhang} and
  {Zhao}}]{yang14false}
\bibinfo{author}{{Yang}, Q.}, \bibinfo{author}{{Yang}, J.},
  \bibinfo{author}{{Yu}, W.}, \bibinfo{author}{{An}, D.},
  \bibinfo{author}{{Zhang}, N.}, \bibinfo{author}{{Zhao}, W.},
  \bibinfo{year}{2014}.
\newblock \bibinfo{title}{On false data-injection attacks against power system
  state estimation: Modeling and countermeasures}.
\newblock \bibinfo{journal}{IEEE Transactions on Parallel and Distributed
  Systems} \bibinfo{volume}{25}, \bibinfo{pages}{717--729}.
\bibitem[{Yang et~al.(2020b)Yang, Liu, Bi and Yang}]{yang20bad}
\bibinfo{author}{Yang, Z.}, \bibinfo{author}{Liu, H.}, \bibinfo{author}{Bi,
  T.}, \bibinfo{author}{Yang, Q.}, \bibinfo{year}{2020}b.
\newblock \bibinfo{title}{Bad data detection algorithm for pmu based on
  spectral clustering}.
\newblock \bibinfo{journal}{Journal of Modern Power Systems and Clean Energy}
  \bibinfo{volume}{8}, \bibinfo{pages}{473--483}.
\bibitem[{Yohanandhan et~al.(2022)Yohanandhan, Elavarasan, Pugazhendhi,
  Premkumar, Mihet-Popa and Terzija}]{yohanandhan2022holistic}
\bibinfo{author}{Yohanandhan, R.V.}, \bibinfo{author}{Elavarasan, R.M.},
  \bibinfo{author}{Pugazhendhi, R.}, \bibinfo{author}{Premkumar, M.},
  \bibinfo{author}{Mihet-Popa, L.}, \bibinfo{author}{Terzija, V.},
  \bibinfo{year}{2022}.
\newblock \bibinfo{title}{A holistic review on cyber-physical power system
  (cpps) testbeds for secure and sustainable electric power grid--part--i:
  Background on cpps and necessity of cpps testbeds}.
\newblock \bibinfo{journal}{International Journal of Electrical Power \& Energy
  Systems} \bibinfo{volume}{136}, \bibinfo{pages}{107718}.
\bibitem[{Zhang and Wang(2006)}]{zhang06detecting}
\bibinfo{author}{Zhang, J.}, \bibinfo{author}{Wang, H.}, \bibinfo{year}{2006}.
\newblock \bibinfo{title}{Detecting outlying subspaces for high-dimensional
  data: the new task, algorithms, and performance}.
\newblock \bibinfo{journal}{Knowledge and information systems}
  \bibinfo{volume}{10}, \bibinfo{pages}{333--355}.
\bibitem[{Zhang et~al.(2020)Zhang, Wang, Weng and Zhang}]{zhang20topology}
\bibinfo{author}{Zhang, J.}, \bibinfo{author}{Wang, Y.}, \bibinfo{author}{Weng,
  Y.}, \bibinfo{author}{Zhang, N.}, \bibinfo{year}{2020}.
\newblock \bibinfo{title}{Topology identification and line parameter estimation
  for non-pmu distribution network: A numerical method}.
\newblock \bibinfo{journal}{IEEE Transactions on Smart Grid} .
\bibitem[{{Zhang} et~al.(2020){Zhang}, {Wang} and {Chen}}]{zhang20detecting}
\bibinfo{author}{{Zhang}, Y.}, \bibinfo{author}{{Wang}, J.},
  \bibinfo{author}{{Chen}, B.}, \bibinfo{year}{2020}.
\newblock \bibinfo{title}{Detecting false data injection attacks in smart
  grids: A semi-supervised deep learning approach}.
\newblock \bibinfo{journal}{IEEE Transactions on Smart Grid} ,
  \bibinfo{pages}{1--1}.
\bibitem[{{Zhao} et~al.(2018){Zhao}, {Zhang}, {Dong} and {La
  Scala}}]{zhao18robust}
\bibinfo{author}{{Zhao}, J.}, \bibinfo{author}{{Zhang}, G.},
  \bibinfo{author}{{Dong}, Z.Y.}, \bibinfo{author}{{La Scala}, M.},
  \bibinfo{year}{2018}.
\newblock \bibinfo{title}{Robust forecasting aided power system state
  estimation considering state correlations}.
\newblock \bibinfo{journal}{IEEE Transactions on Smart Grid}
  \bibinfo{volume}{9}, \bibinfo{pages}{2658--2666}.
\bibitem[{Zhao et~al.(2015)Zhao, Zhang, Dong and Wong}]{zhao15forecasting}
\bibinfo{author}{Zhao, J.}, \bibinfo{author}{Zhang, G.}, \bibinfo{author}{Dong,
  Z.Y.}, \bibinfo{author}{Wong, K.P.}, \bibinfo{year}{2015}.
\newblock \bibinfo{title}{Forecasting-aided imperfect false data injection
  attacks against power system nonlinear state estimation}.
\newblock \bibinfo{journal}{IEEE Transactions on Smart Grid}
  \bibinfo{volume}{7}, \bibinfo{pages}{6--8}.
\bibitem[{{Zhao} et~al.(2017){Zhao}, {Zhang}, {La Scala}, {Dong}, {Chen} and
  {Wang}}]{zhao15short}
\bibinfo{author}{{Zhao}, J.}, \bibinfo{author}{{Zhang}, G.},
  \bibinfo{author}{{La Scala}, M.}, \bibinfo{author}{{Dong}, Z.Y.},
  \bibinfo{author}{{Chen}, C.}, \bibinfo{author}{{Wang}, J.},
  \bibinfo{year}{2017}.
\newblock \bibinfo{title}{Short-term state forecasting-aided method for
  detection of smart grid general false data injection attacks}.
\newblock \bibinfo{journal}{IEEE Transactions on Smart Grid}
  \bibinfo{volume}{8}, \bibinfo{pages}{1580--1590}.
\bibitem[{{Zimmerman} et~al.(2011){Zimmerman}, {Murillo-Sánchez} and
  {Thomas}}]{zimmerman11matpower}
\bibinfo{author}{{Zimmerman}, R.D.}, \bibinfo{author}{{Murillo-Sánchez},
  C.E.}, \bibinfo{author}{{Thomas}, R.J.}, \bibinfo{year}{2011}.
\newblock \bibinfo{title}{Matpower: Steady-state operations, planning, and
  analysis tools for power systems research and education}.
\newblock \bibinfo{journal}{IEEE Transactions on Power Systems}
  \bibinfo{volume}{26}, \bibinfo{pages}{12--19}.

\end{thebibliography}

\end{document}